\documentclass{egpubl}
\usepackage{mathptmx}
\usepackage{graphicx}

\usepackage{color}
\usepackage{times}
\usepackage[ruled]{algorithm2e} 

\usepackage{bm}
\SetAlFnt{\small}
\SetAlCapFnt{\small}
\SetAlCapNameFnt{\small}
\SetAlCapHSkip{0pt}
\usepackage{verbatim}
\usepackage{subfigure}
\usepackage{amsmath}
\usepackage{amsfonts}
\usepackage{overpic}

\JournalSubmission    
\usepackage[T1]{fontenc}
\usepackage{dfadobe}  

\biberVersion
\BibtexOrBiblatex
\electronicVersion
\PrintedOrElectronic

\usepackage{egweblnk} 

\usepackage{color,overpic}

\newcommand{\cbl}[1]{{\color{blue}{#1}}}

  \title {Turbulent Details Simulation\\ for SPH Fluids via Vorticity Refinement 
  }

 \author[S. Liu \& X. Wang \& X. Ban et al.]
 {\parbox{\textwidth}{\centering Sinuo Liu$^{2}$\thanks{Co-first author, contributed equally}, Xiaokun Wang$^{1}\footnotemark[1]$\thanks{Corresponding author: banxj@ustb.edu.cn; wangxiaokun@ustb.edu.cn},   
         Xiaojuan Ban$^{1}$\footnotemark[2], Yanrui Xu$^{1}$, Jing Zhou$^{2}$,
      Ji\v{r}\'i Kosinka$^3$,  Alexandru C. Telea$^4$
        }
      \\
 {\parbox{\textwidth}{\centering $^1$University of Science \& Technology Beijing, Institute of Artificial Intelligence, China\\
  $^2$University of Science \& Technology Beijing, School of Computer \& Communication Engineering, China\\
         $^3$University of Groningen, Bernoulli Institute, Netherlands
         \\
         $^4$Utrecht University, Department of Information and Computing Sciences, Netherlands
      }
 }
}

\begin{document}

\teaser{
\centering
\begin{overpic}[width=16cm]{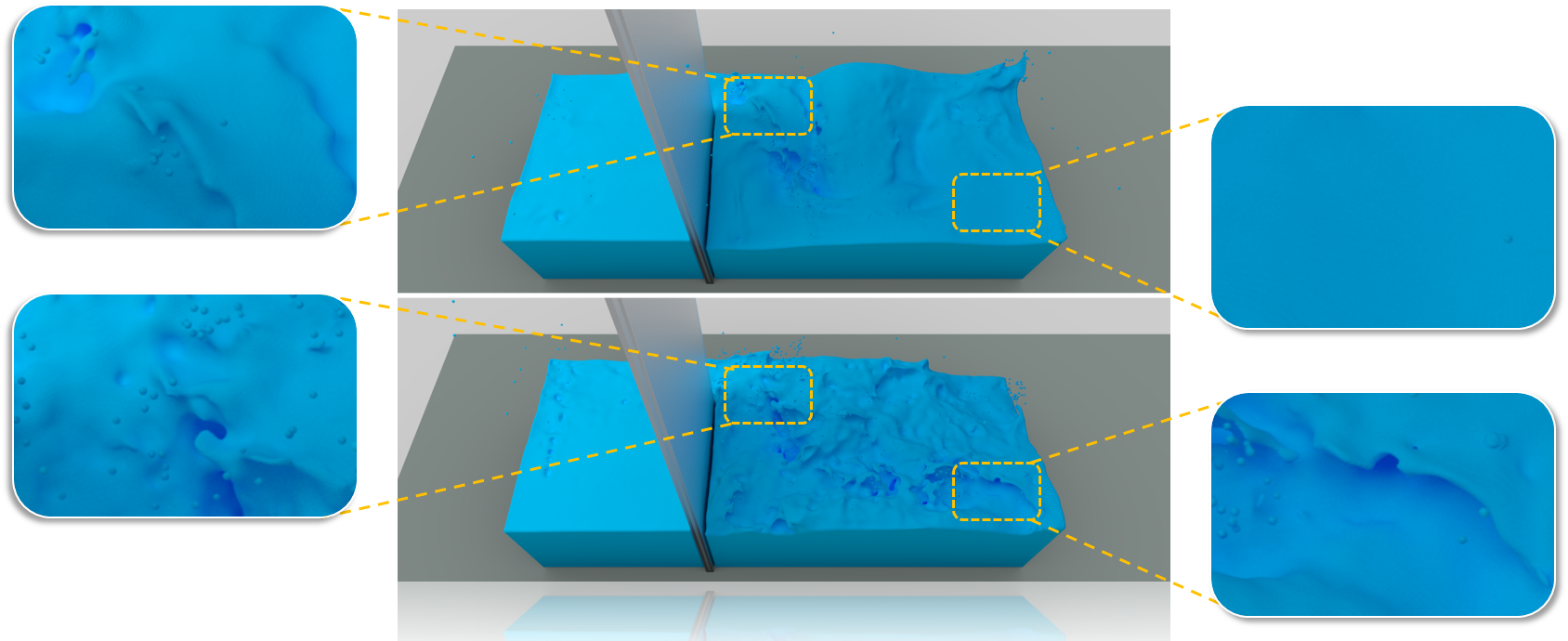}
 \put(45,20){\footnotesize With vorticity refinement}
 \put(45,38.4){\footnotesize Without vorticity refinement}
 \end{overpic}
  \caption{Our Vorticity Refinement (VR) solver applied to an DFSPH \cite{bender2015divergence} simulation (1.18M particles). In this scene, a breaking dam collides with a board, creating  turbulence. Zoom-ins compare the surface under DFSPH without (top) and with (bottom) our VR solver. The result shows that our method better captures turbulence details.}
  \label{fig:abs}
}

\maketitle
\begin{abstract}
   A major issue in Smoothed Particle Hydrodynamics (SPH) approaches is the numerical dissipation during the projection process, especially under coarse discretizations. High-frequency details, such as turbulence and vortices, are smoothed out, leading to unrealistic results. To address this issue, we introduce a Vorticity Refinement (VR) solver for SPH fluids with negligible computational overhead. In this method, the numerical dissipation of the vorticity field is recovered by the difference between the theoretical and the actual vorticity, so as to enhance turbulence details. Instead of solving the Biot-Savart integrals, a stream function, which is easier and more efficient to solve, is used to relate the vorticity field to the velocity field. We obtain turbulence effects of different intensity levels by changing an adjustable parameter. Since the vorticity field is enhanced according to the curl field, our method can not only amplify existing vortices, but also capture additional turbulence. Our VR solver is straightforward to implement and can be easily integrated into existing SPH methods.


\end{abstract}  

\section{Introduction}

Fluid simulation is a hot topic in computer graphics, with huge research and application demands. Within this context, the Smoothed Particle Hydrodynamics (SPH) method simulates fluids with large deformations accurately and efficiently, showing abundant details and vivid motion. In the past decades, several solutions have been proposed to enforce incompressibility \cite{solenthaler2009predictive, ihmsen2014implicit,bender2015divergence}. However, numerical dissipation problems still remain and cause a significant loss of turbulence details \cite{jiang2017angular,Fu2017PPM}. For instance, vorticity dissipation is one of the major issues causing the loss of details on the fluid surface and in overall dynamic effects \cite{koschier2019smoothed}.

To maintain complex turbulence and vortex details on the fluid surface, some methods proposed to increase the apparent resolution by seeding over surface points~\cite{mercier2015surface}, or use an adaptive volumetric mesh for grid-based fluids~\cite{edwards2014detailed}. However, these methods only add details over a coarse discretization, without considering the inner volume. Vorticity Confinement (VC) methods add vortices from the perspective of the entire flow field \cite{fedkiw2001visual,macklin2013position} to recover dissipated details. However, VC methods tend to add more energy than is dissipated, and can amplify only existing vortices. Lagrangian vortex methods, such as vortex particles\,\cite{park2005vortex} and vortex filaments\,\cite{weissmann2010filament}, have been used to effectively simulate turbulent fluids. While these methods maintain a divergence-free velocity field and have theoretically no numerical dissipation, they require solving the equivalent of three Poisson equations to obtain velocity from vorticity, which is computationally expensive.

To alleviate the above-mentioned problems and obtain more realistic turbulent flows, we introduce a turbulence refinement scheme by correcting the vorticity field. In continuum mechanics, vorticity is a pseudovector field that describes the local spinning motion of a continuum. It can be defined as the curl of the fluid’s velocity field. Like the divergence error issue mentioned in the DFSPH method\,\cite{bender2015divergence}, vorticity dissipation also reduces the realism of simulations. To date, 
the kinetic energy from the vorticity field could be transformed into positive divergence, causing the loss of surface details and of overall dynamic motion\,\cite{zhang2015restoring}. State-of-the-art SPH approaches for fluid simulation cannot solve this problem completely. 


\begin{figure}[t]
    \centering
    \includegraphics[scale=0.23]{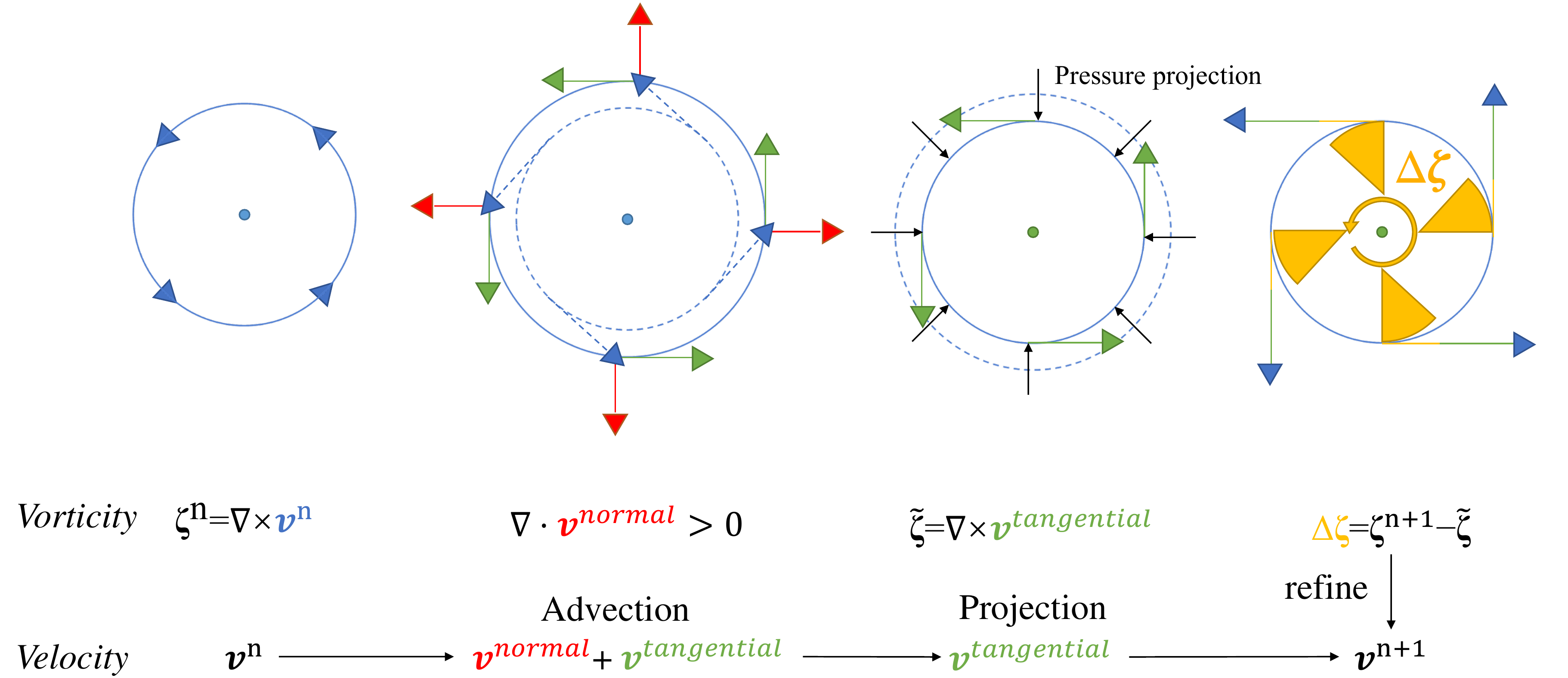}
    \caption{This diagram shows how the vorticity dissipates in the advection-projection process. Linear velocity at time step $n$ splits into normal (red) and tangential (green) components during the advection procedure. Next, projection eliminates the normal component using the pressure force to keep the flow divergence-free.}
    \label{fig:my_label}
\end{figure}

\begin{figure*}[htb]
\centering  
\subfigcapskip=2pt 
\subfigbottomskip=0pt
\subfigtopskip=0pt

\subfigure[DFSPH]{
\begin{minipage}[c]{0.32\linewidth}
\includegraphics[scale=0.18]{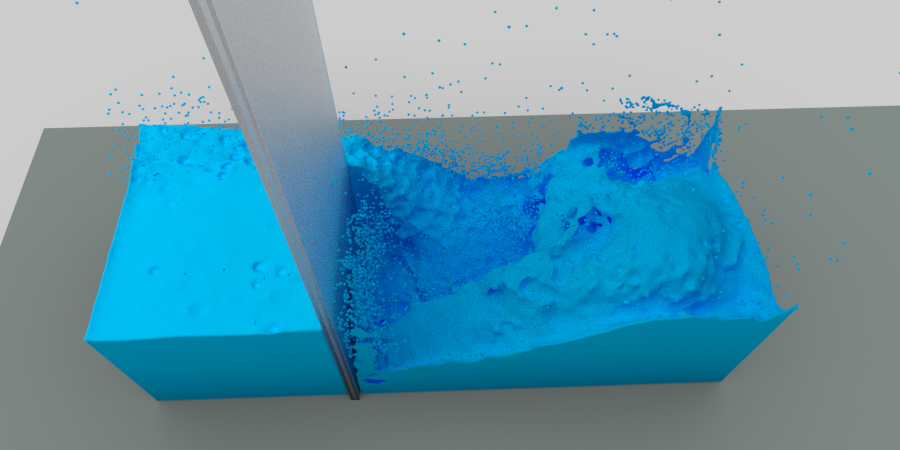}
\end{minipage}%
\begin{minipage}[c]{0.32\linewidth}
\includegraphics[scale=0.18]{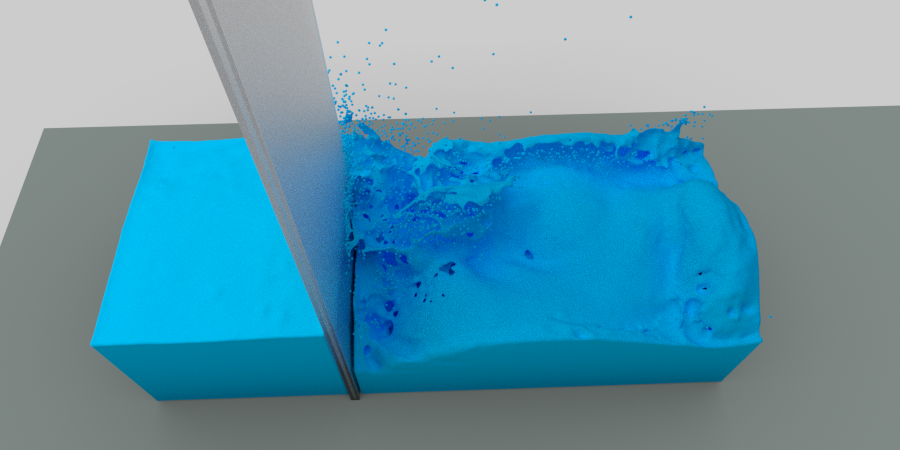}
\end{minipage}\!
\begin{minipage}[c]{0.32\linewidth}
\includegraphics[scale=0.18]{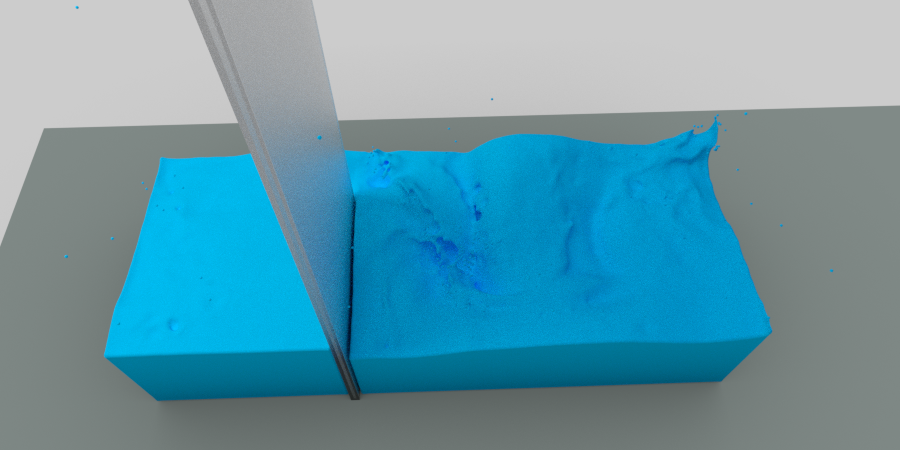}
\end{minipage}
}

\subfigure[MP solver]{
\begin{minipage}[c]{0.32\linewidth}
\includegraphics[scale=0.18]{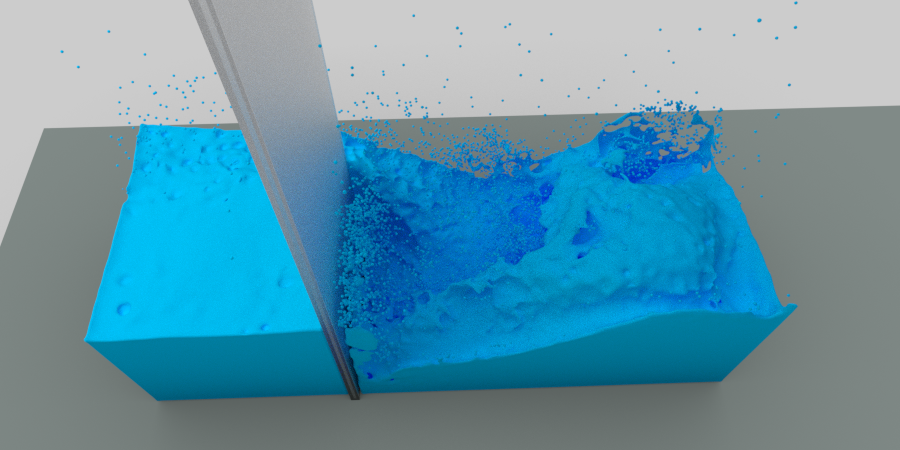}
\end{minipage}%
\begin{minipage}[c]{0.32\linewidth}
\includegraphics[scale=0.18]{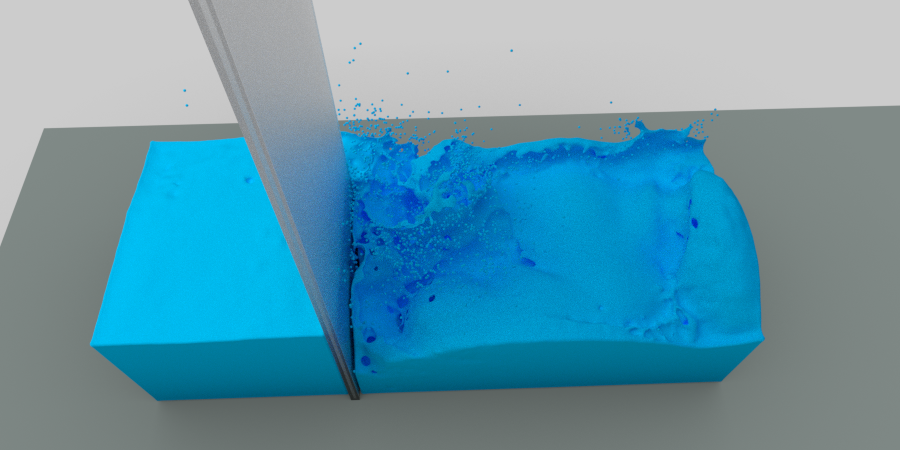}
\end{minipage}\!
\begin{minipage}[c]{0.32\linewidth}
\includegraphics[scale=0.18]{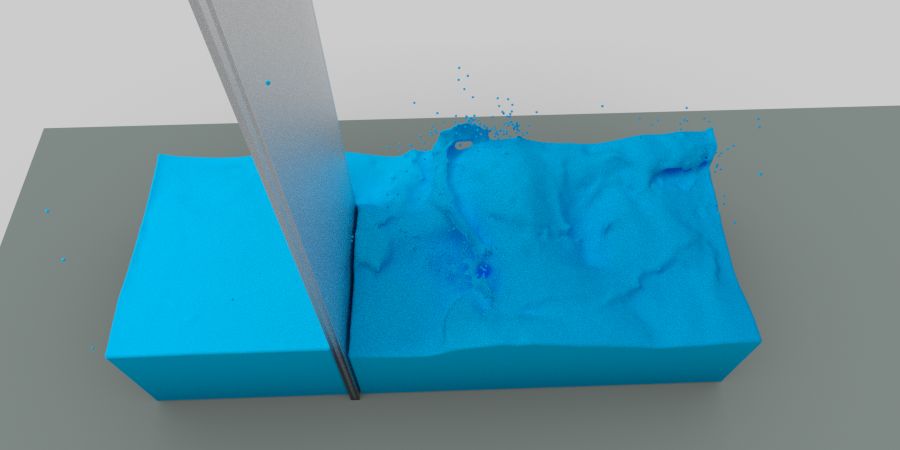}
\end{minipage}
}

\subfigure[Our method]{
\begin{minipage}[c]{0.32\linewidth}
\includegraphics[scale=0.18]{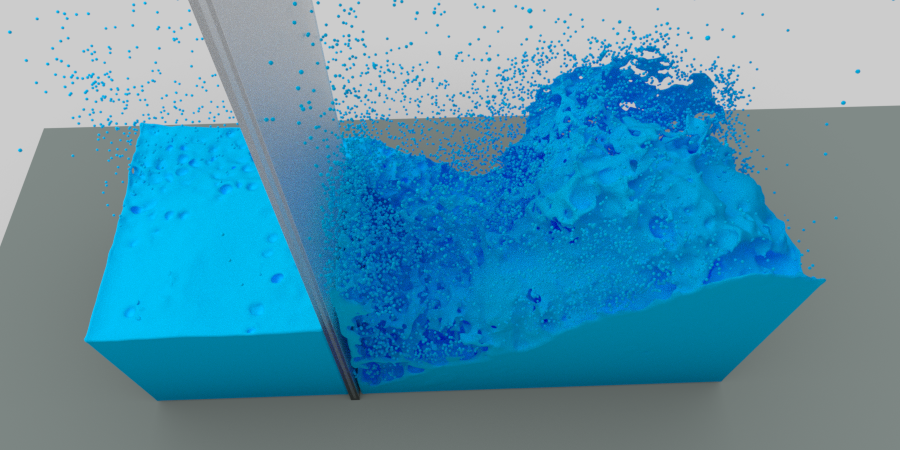}
\end{minipage}%
\begin{minipage}[c]{0.32\linewidth}
\includegraphics[scale=0.18]{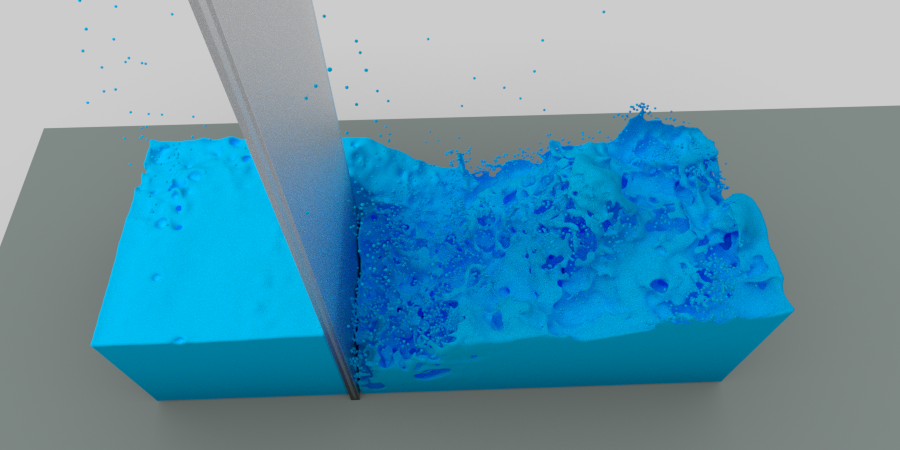}
\end{minipage}\!
\begin{minipage}[c]{0.32\linewidth}
\includegraphics[scale=0.18]{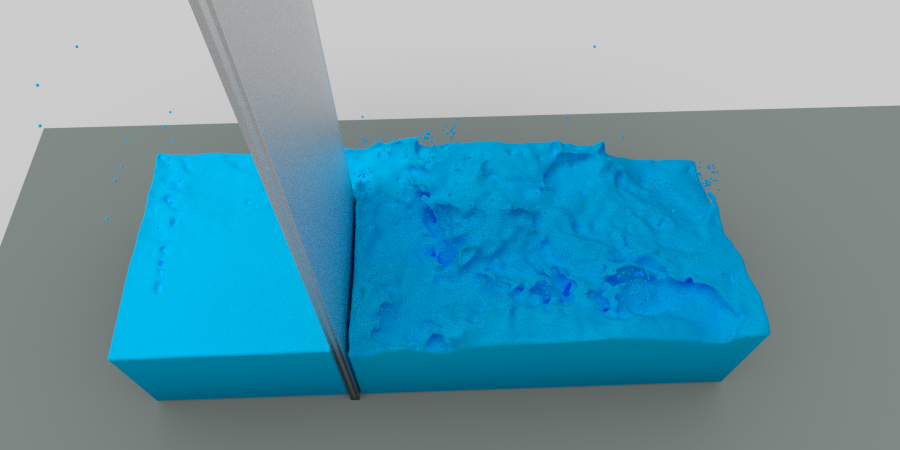}
\end{minipage}
}

\caption{A breaking dam collides with a board (1.18M particles). (a) Only few vortex effects are formed using DFSPH. (b) The MP solver enhances the simulation result to some extent. (c) Our method greatly improves the turbulence details, and the surface details are richer than with the MP solver. The MP solver and our solver achieve different styles.}
  \label{fig:board} 
\end{figure*}

During the advection-projection process, the advection step maps the original velocity field into a rotational part and a divergent part, after which the pressure projection removes the divergent part, leaving only the rotational part. The angular momentum is therefore lost in the simulation, with the effect becoming worse as the time step size increases; see Fig.~\ref{fig:my_label}, the orange part of the diagram. To alleviate this, we use the accurate vorticity field derived from the curl of the Navier-Stokes equations to correct the linear velocity for each particle; see Fig.~\ref{fig:my_label}, the green part of the diagram. Moreover, we use a stream function to refine the velocity using a reasonable augmentation of the vorticity field which can restore vivid yet controllable vortices and turbulence effects (as shown in Figs.\@~\ref{fig:abs} and~\ref{fig:board} among others). Previous related work\,\cite{xu2019turbulence} looked into correcting the velocity field through the vorticity recovered from the kinematic viscosity by increasing the vorticity field proportionally by the energy dissipated, which is based on the rotational kinetic energy. In contrast, we focus on getting the ideal vorticity field directly from the curl of the Navier-Stokes equations, which is a more physically reasonable model.

Summarizing, the main contributions of this paper are:
\begin{itemize}
\item A vivid turbulence-details generation method that \cbl{recovers} numerical dissipation through vorticity field correction;
\item A novel vorticity-based constraint and stream function solution for simulating turbulence;
\item An orthogonal solver for the SPH fluid framework with turbulence simulation that can be easily integrated into other particle-based methods and fluid solvers.
\end{itemize}

The remainder of the paper is organized as follows. Section~2 gives an overview of related work. Section~3 discusses the accuracy of numerical calculations in SPH. Our vorticity refinement scheme is presented in Section~4.
Section~5 presents and discusses our experimental results. Finally, Section~6 concludes the paper.

\section{Related Work}

Fluid simulation is a well researched topic in computer graphics. Early works on this topic include\,\cite{monaghan1994simulating,foster1996realistic,stam1999stable,muller2003particle}. For recent overviews, we refer to Bridson's book~\cite{bridson2015fluid} and the state-of-the-art report of Koschier et al.~\cite{koschier2019smoothed}.
We further discuss more specific work related to our context, namely SPH-based fluid simulation (Sec.~2.1) and turbulence simulation (Sec.~2.2).

\subsection{Incompressibility and Numerical Accuracy in SPH Fluid Simulations}
Monaghan simulated free surface flows with SPH~\cite{monaghan1994simulating}, which laid the foundation for fluid simulation. Later, Muller et al.~\cite{muller2003particle} proposed to simulate fluids using the ideal gas state equation with surface tension and viscosity forces, but without full incompressibility. An improved weakly-compressible SPH (WCSPH) method was proposed by Becker and Teschner~\cite{becker2007weakly}. The use of the stiff equation of state (EOS) significantly increased realistic effects, but the efficiency of such methods is limited by the size of the used time step. To further enforce incompressibility and improve numerical accuracy, much effort has been invested into implicit pressure solvers. Previous approaches can be categorized as methods that project particle positions onto an incompressible state using iterative EOS solvers, and pressure projection methods~\cite{ihmsen2014sph}, as follows.

He et al.~\cite{he2012local} and Solenthaler and Pajarola~\cite{solenthaler2009predictive} proposed predictive-corrective approaches that iteratively project particle positions onto an incompressible state. This is also done in position-based fluids (PBF)~\cite{macklin2013position}. However, PBF avoids accumulating pressure or pressure forces that eventually update the velocity and the position. Ihmsen et al.~\cite{ihmsen2014implicit} proposed implicit incompressible SPH (IISPH) following the strategy of pressure projections. Separately, Bender and Koschier~\cite{bender2015divergence} proposed a method that enforces a low compression ($0.01\%$ ) and a divergence-free velocity constraint (DFSPH). Among all the variants of the SPH method, the typical advection-projection models are PCISPH~\cite{solenthaler2009predictive}, IISPH~\cite{ihmsen2014implicit} and DFSPH~\cite{bender2015divergence}. In this paper, we use the DFSPH approach as a baseline for comparisons of computational efficiency and stability.

It is well known that SPH approaches suffer from  numerical dissipation problems, especially for coarse discretizations~\cite{monaghan1994simulating, de2015power, bender2018turbulent}. Ihmsen et al.~\cite{ihmsen2014sph} pointed out that SPH results in undesired dissipation and high-frequency features are smoothed out. Therefore, avoiding dissipation for turbulence in fluid simulation is needed to improve visual realism. 

\subsection{Restoring Turbulence in Fluid Simulation}

Restoring high-frequency details has been an important challenge in fluid simulation since its very beginning~\cite{kim2008wavelet,Jeschke2018WSW}. For Eulerian approaches, Stam's scheme~\cite{stam1999stable} first achieved realistic and real-time fluid simulation on consumer-grade graphics hardware. However, the first-order accuracy in both time and space makes this method (and other extensions thereof) suffer from serious numerical dissipation. Kim et al.~\cite{kim2005flowfixer} proposed a higher-order approximation. Jonas et al.~\cite{zehnder2018advection} proposed an advection-reflection solver for detail-preserving fluid animation which leads to two orders of magnitude reduction in energy loss. Rahul et al.~\cite{narain2019second} then established a connection between this method and the implicit midpoint time integration scheme, and presented a simple improvement to obtain an advection-reflection scheme with second-order accuracy in time.

Hybrid particle-grid methods were subsequently proposed to further reduce numerical dissipation. Zhu and Bridsons’ FLIP method for incompressible flow~\cite{zhu2005animating} significantly eliminates the dissipation in advection. Jiang et al.~\cite{jiang2015affine} successfully restore most of the rotational motion using a hybrid method.  

Although the general simulation methods mentioned above can handle numerical dissipation on a macroscopic level, both Eulerian and Lagrangian approaches face challenges when simulating high-frequency details such as turbulence. Therefore, methods specifically designed for refining turbulent details have emerged. These can be classified into three categories: up-res methods, vorticity confinement methods, and Lagrangian vortex methods \cite{bender2018turbulent}, as follows.

\noindent\textbf{Up-res methods} add high-frequency details over a coarse discretization. Mercier et al.~\cite{mercier2015surface} proposed a  post-processing method to apply fine turbulence over particle-based fluid surfaces. High-resolution surface points are seeded after curvature evaluation, and the detailed surface waves are then evolved over coarse particles. Edwards and Bridson~\cite{edwards2014detailed} proposed an adaptive volumetric-mesh method for grid-based fluids by using the adaptive discontinuous Galerkin method. Machine learning methods such as Convolutional Neural Networks (CNNs)~\cite{chu2017data} have been applied in fluid simulation to synthesize high-resolution turbulence on rough simulation results based on a high-resolution source. However, training CNNs is time-consuming and often requires delicate hyperparameter tuning. Overall, up-res methods can typically improve only surface effects.

\noindent\textbf{Vorticity confinement methods} aim to find existing vortices and recover their dissipation. A new forcing term is added to increase the velocity of target positions, and to enforce the rotation, of the vortex. Lentine et al.~\cite{lentine2011mass} improved vorticity confinement to be both energy conserving and momentum conserving. Jang et al.~\cite{jang2010multilevel} used multi-level vorticity confinement to acquire better results. Macklin and Muller~\cite{macklin2013position} presented a simple method to amplify the existing vorticity through accelerating particles using SPH. Overall, vorticity confinement methods provide a simple way for preserving vortices, but are in general unable to create additional turbulence details. Moreover, they are prone to adding excessive energy to the system so that energy conservation is likely to be violated, leading to unstable results.

\noindent\textbf{Lagrangian vortex methods} build on the vorticity representation of the Navier-Stokes equations~\cite{park2005vortex}, which have less numerical dissipation and more divergence retention than vorticity confinement methods. These methods can be applied to particles~\cite{wang2020robust}, curves~\cite{angelidis2005simulation}, filaments~\cite{eberhardt2017hierarchical}, and even surfaces~\cite{weissmann2010filament}. 
Yet, boundaries, such as non-rigid obstacles and free surfaces, are difficult to handle. Zhu et al.~\cite{zhu2010creating} proposed to simulate vortex details around moving objects using Eulerian grids. Golas et al.~\cite{golas2012large} also treated boundaries of an Eulerian grid to solve this issue. A disadvantage of these methods is that the velocity field has to be recovered by solving the Biot-Savart integrals or a vector-valued Poisson equation. Recently, Bender et al.~\cite{bender2018turbulent} introduced the MicroPolar fluid solver (MP solver) for inviscid fluids in order to capture the micro-rotation of fluid particles, achieving impressive visual turbulent features. Wang et al.~\cite{wang2020robust} proposed a turbulence refinement method based on the Rankine vortex model for particle-based simulation. Zhang et al.~\cite{zhang2015restoring} proposed an Integrated Vorticity of Convective Kinematics (IVOCK) method to restore dissipated energy by measuring vorticity loss in advection. This method can cheaply capture much of the lost details for smoke and fire, but does not work well for liquid simulations. In~\cite{zhang2015restoring}, only the vorticity dissipated during the advection step is considered. The refined linear velocity in their paper is the velocity after the advection step. This velocity is then further affected by viscosity and the projection step. Viscosity may become another source of vorticity dissipation and the pressure force may introduce vorticity errors into the velocity field after the projection step. Although it maintains an incompressible density field, it is not necessarily divergence-free.

\noindent\textbf{Our method} is inspired by the idea of stream functions~\cite{zhang2015restoring}, extended to Lagrangian fluid simulations. This allows us to efficiently derive velocity refinement from the vorticity field. Recovering turbulence from the curl form of the Navier-Stokes equations has a long history. In 2005, Park and Kim~\cite{park2005vortex} gave the governing equations of the vortex method and introduced the concept of the stream function. \cite{zhang2015restoring} and our work, among many other vortex methods, utilize this concept to reduce numerical dissipation during simulation. In our method, we derive the dissipated vorticity during the whole advection-projection step in the SPH approach. This can be easily done with little extra computation overhead. With respect to the concept of the stream function, we carry out the Biot-Savart summation process within smoothing length, which makes it less accurate but more efficient than~\cite{zhang2015restoring}; we show this to be sufficient to maintain stability. This is because, theoretically, the refined velocity is the curl of the stream function, and any curl of a vector field is divergence-free. Moreover, we implemented our method using DFSPH (Divergence-free SPH), which includes an extra divergence-correction solver, thereby eliminating possible errors caused by the summation process. Moreover, we do not need to solve the Biot-Savart integrals or a vector-valued Poisson equation. In contrast to the MP solver~\cite{bender2018turbulent}, in which the motion equation is obtained from the MicroPolar model and discretized with SPH, we derive the vorticity equation from the curl of the Navier-Stokes equations, and recover velocity from the vorticity field using stream functions. Our results show that our method can not only enhance existing vortices, but also generate turbulence at potential locations of new vortices.

\begin{figure*}[tbp]
\centering
\begin{tabular}{c|c|c|c}

\includegraphics[scale=0.37]{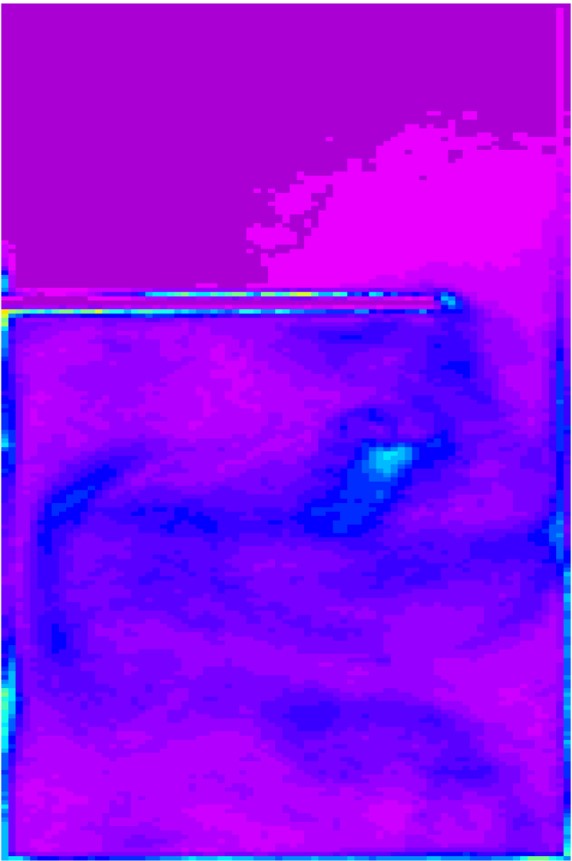} & 
\includegraphics[scale=0.37]{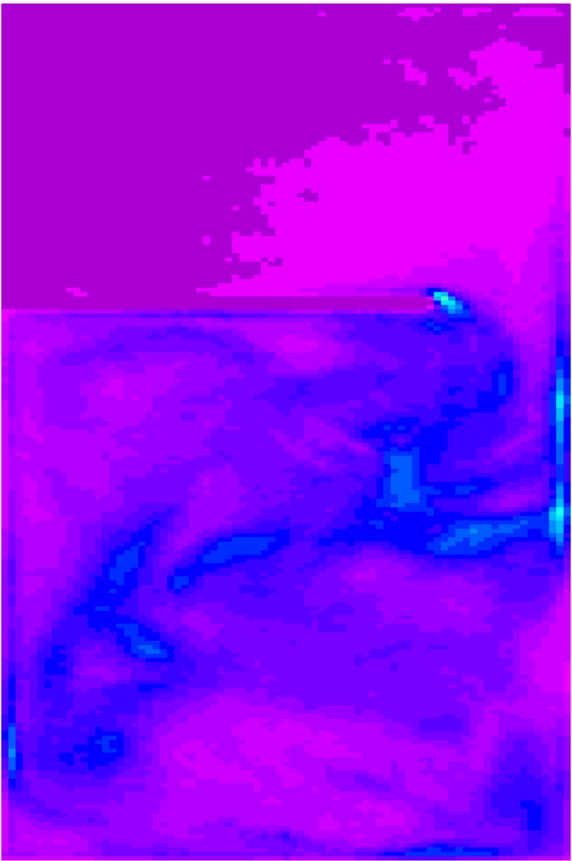} & 
\includegraphics[scale=0.37]{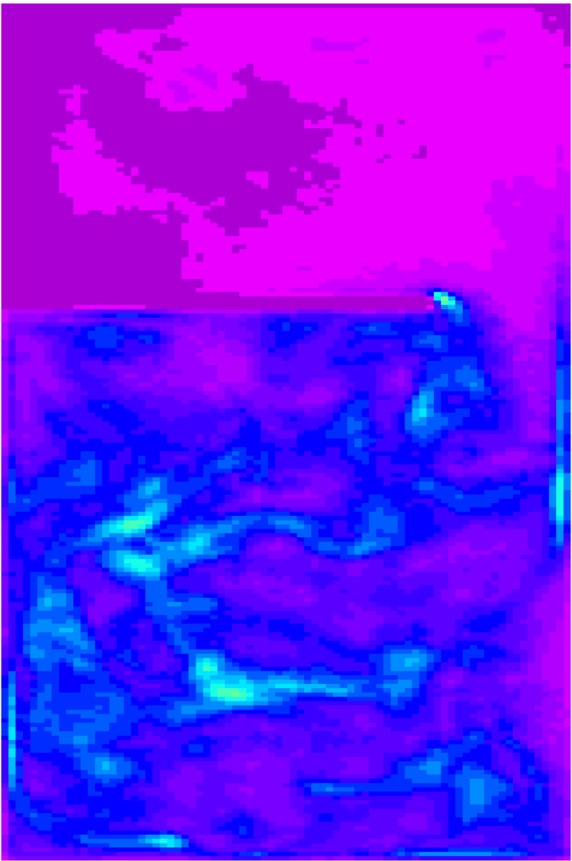}  &
\includegraphics[scale=0.37]{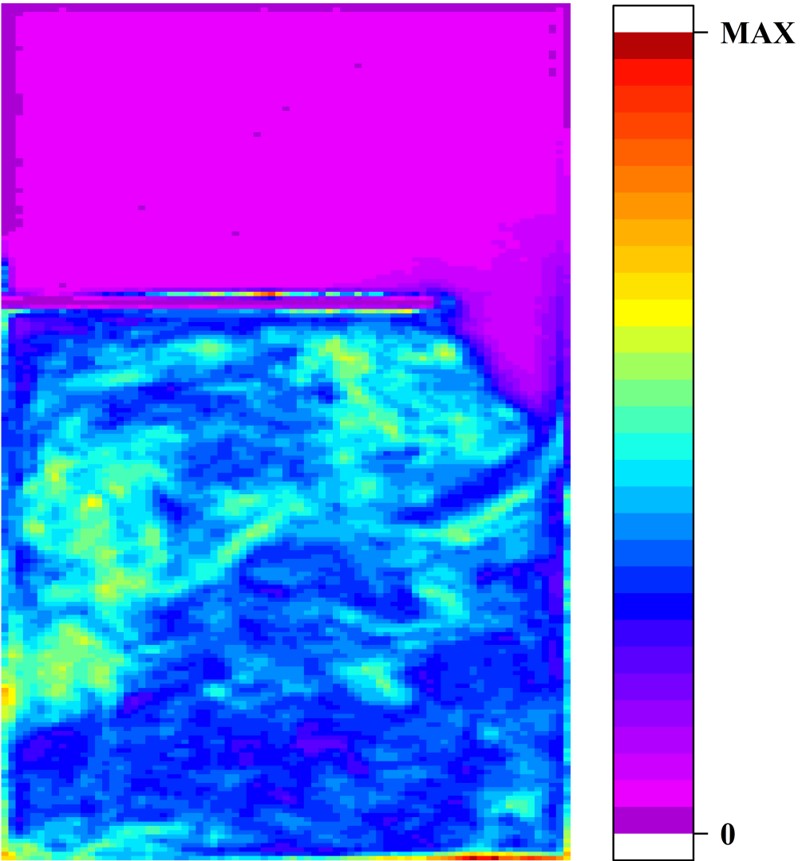}\\
DFSPH & MP solver ($\nu_t=0.05$) & MP solver ($\nu_t$=0.2)&Our method ($\alpha = 1.0$)
\end{tabular}
\caption{Comparison of vorticity in a 2D scene extracted from the 3D scene in Fig.\@~\ref{fig:board}. This visualization corresponds to the second column in Fig.~\ref{fig:board}. Color shows the vorticity magnitude of the particles, thereby allowing one to compare the vorticity of DFSPH, the MP solver, and our VR method.}
\label{fig.particle vorticity}
\end{figure*}

\section{SPH Discretization for Fluid Simulation}
\label{sec:sim}

Traditional Lagrangian-based fluid simulations use the fluid governing equations, the Navier-Stokes equations, to solve for the position and velocity of each fluid particle. The acceleration of the fluid particles is obtained by the combination of pressure $\mathbf{a}_{pres}$, viscous force $\mathbf{a}_{vis}$, and gravity $\mathbf{a}_{g}$ as
\begin{equation}
\label{eqn:01-1}
\frac{D\bm{v}}{Dt} = \mathbf{a}_{pres} + \mathbf{a}_{vis} +\mathbf{a}_{g} = -\frac{1}{\rho}\nabla p + \nu_v\nabla^{2}\bm{v} + \mathbf{g},
\end{equation}
where $D$ denotes the material derivative, $\rho$ is the density of the fluid, $p$ represents pressure, $\bm{v}$ is velocity, $\nu_v$ is the kinematic viscosity coefficient, a value that characterizes various fluid types (set to $\nu_v=0.05$ in our experiments), $\mathbf{g}$ is the gravitational acceleration, and $\nabla^2$ denotes the Laplace operator.

The SPH approach can be used to discretize the Navier-Stokes equations to numerically solve them. The continuous physical values in space can be discretized using a smooth kernel $W$ as in
\begin{equation}
\label{ssph}
    \mathbf{A}\left( {{\bm{x}_i}} \right) = \sum\limits_{\|\bm{x}_i-\bm{x}_j\| \leq h} {{m(\bm{x}_j)}\frac{{{\mathbf{A}(\bm{x}_j)}}}{{{\rho(\bm{x}_j})}}} W\left( {{\bm{x}_i} - {\bm{x}_j},h} \right)
\end{equation}
with $\mathbf{A}\left( {{\bm{x}_i}} \right)$ being a certain quantity associated with particle $i$ at location ${\bm{x}_i}$. This quantity can be interpolated from the values of neighbour particles, indexed by $j$, within a support radius $h$. 
The quantities $m$ and $\rho$ stand for mass and density, respectively.
To simplify notation, we next use the shorthand $\mathbf{A}_i$ to denote the quantity $\mathbf{A}$ evaluated at position $\bm{x}_i$.

The density of a fluid can be derived by simply replacing $\mathbf{A}$ by $\rho$. In our work, we use the \emph{cubic spline kernel}~\cite{monaghan1985particle}:
\begin{equation}\nonumber
W_{ij} = \frac{1}{\pi h^3}
\left\{\begin{matrix}
 1-\frac{2}{3}x^2 + \frac{3}{4}x^3 & 0 \le x \le 1 \\[1mm]
 \frac{1}{4} \left ( 2-x \right)^3 & 1 \le x \le 2 \\[1mm]
 0 & x \geq 2
\end{matrix}\right. ,
\end{equation}
where $x=\| {\bm{x}_{i}-\bm{x}_{j}} \| / h$ and $W_{ij}$ is a short form of $W\!\left( {{\bm{x}_i} - {\bm{x}_j},h} \right)$.
To obtain a better accuracy of the approximation of the divergence of velocity, the gradient and the curl of velocity, we apply the difference form of the SPH discretization as:
\begin{equation} \label{diffSPH}
\nabla \otimes \mathbf{A} = \sum_j{
\frac{m_j}{\rho_j}
\left (
\mathbf{A}_j-\mathbf{A}_i
\right )
\otimes \nabla W_{ij}
},
\end{equation}
which expresses the gradient ($\nabla A$), divergence ($\nabla\cdot A$), and curl ($\nabla\times A$, in which case the right hand side is negative) of $A$.
Since the second derivative is often sensitive to particle disorder and sign changes inside the support radius $h$, we use artificial viscosity to approximate the Laplacian as follows \cite{koschier2019smoothed}:

\begin{equation}
\label{eqn:04_1}
\nabla^{2}\mathbf{A}(\bm{x}_i) = 2(d+2) \sum_{j} \frac{m_{j}}{\rho_{j}} \frac{\mathbf{A}_{ij} \cdot \bm{x}_{ij} }{{ {\bm{x}_{ij}} }^2 + 0.01 h^2} \nabla W_{ij},
\end{equation}
where $d$ is the space dimension (in our case equal to $3$), $\bm{x}_{ij} = \bm{x}_i-\bm{x}_j$, and $\mathbf{A}_{ij} = \mathbf{A}_i - \mathbf{A}_j$. 



\begin{figure}[htb]
\centering
\subfigcapskip=2pt 
\subfigbottomskip=0pt
\subfigtopskip=0pt
\subfigure[DFSPH]{
\begin{minipage}[c]{1.0\linewidth}
\includegraphics[scale=0.207]{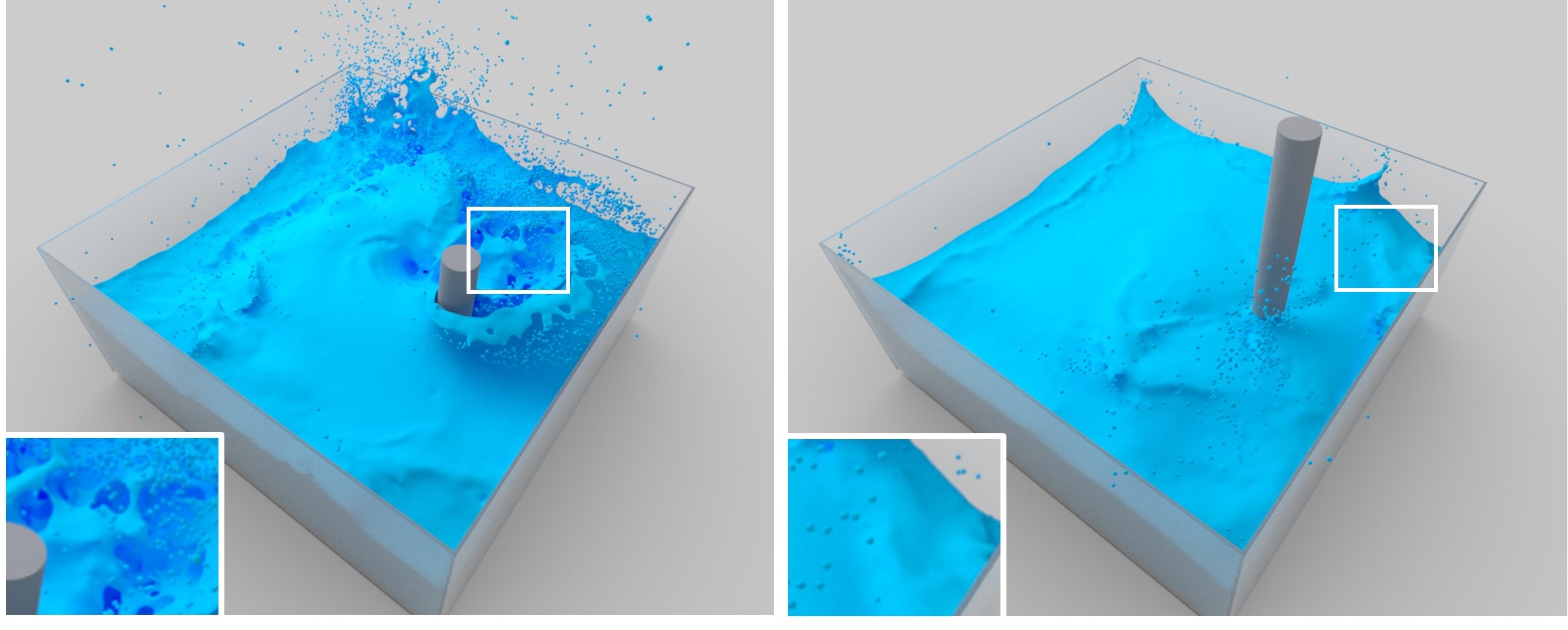}
\end{minipage}
}
\subfigure[MP solver ($\nu_t=0.05$)]{
\begin{minipage}[c]{1.0\linewidth}
\includegraphics[scale=0.207]{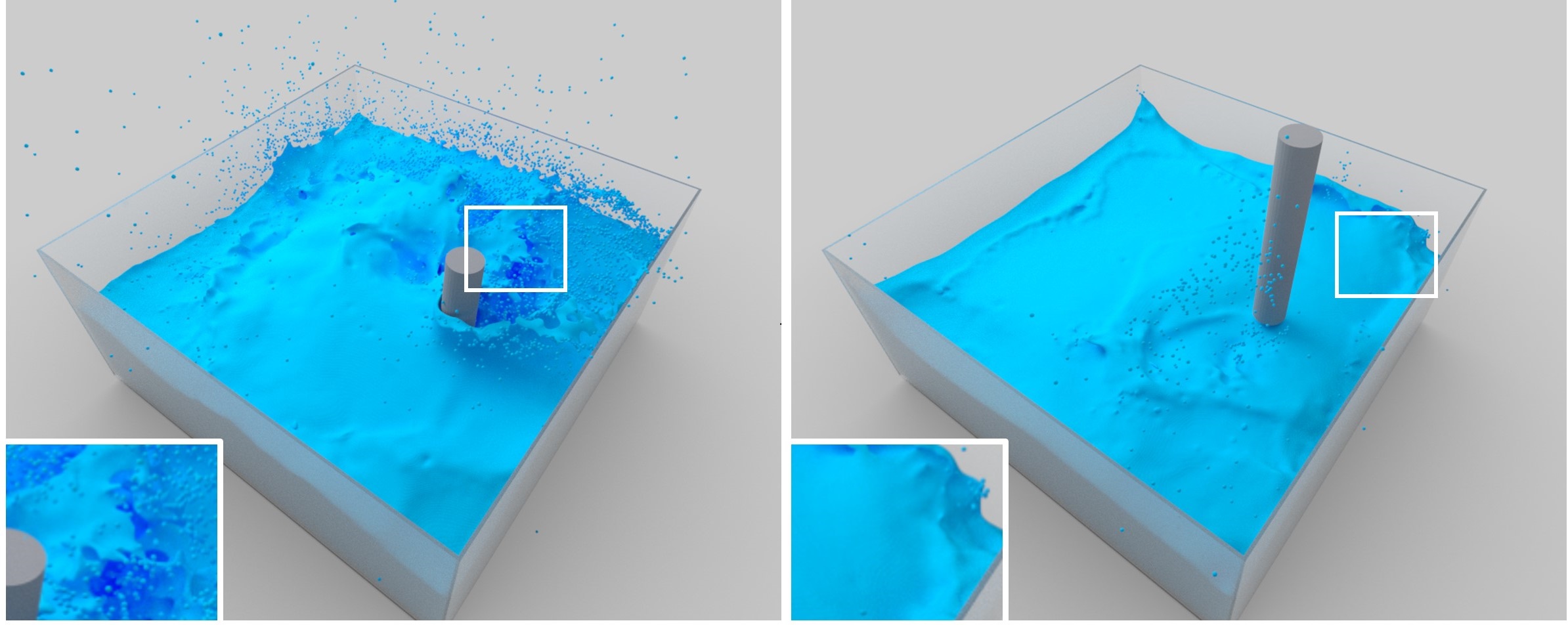}
\end{minipage}
}
\subfigure[Our method ($\alpha=1.0$)]{
\begin{minipage}[c]{1.0\linewidth}
\includegraphics[scale=0.207]{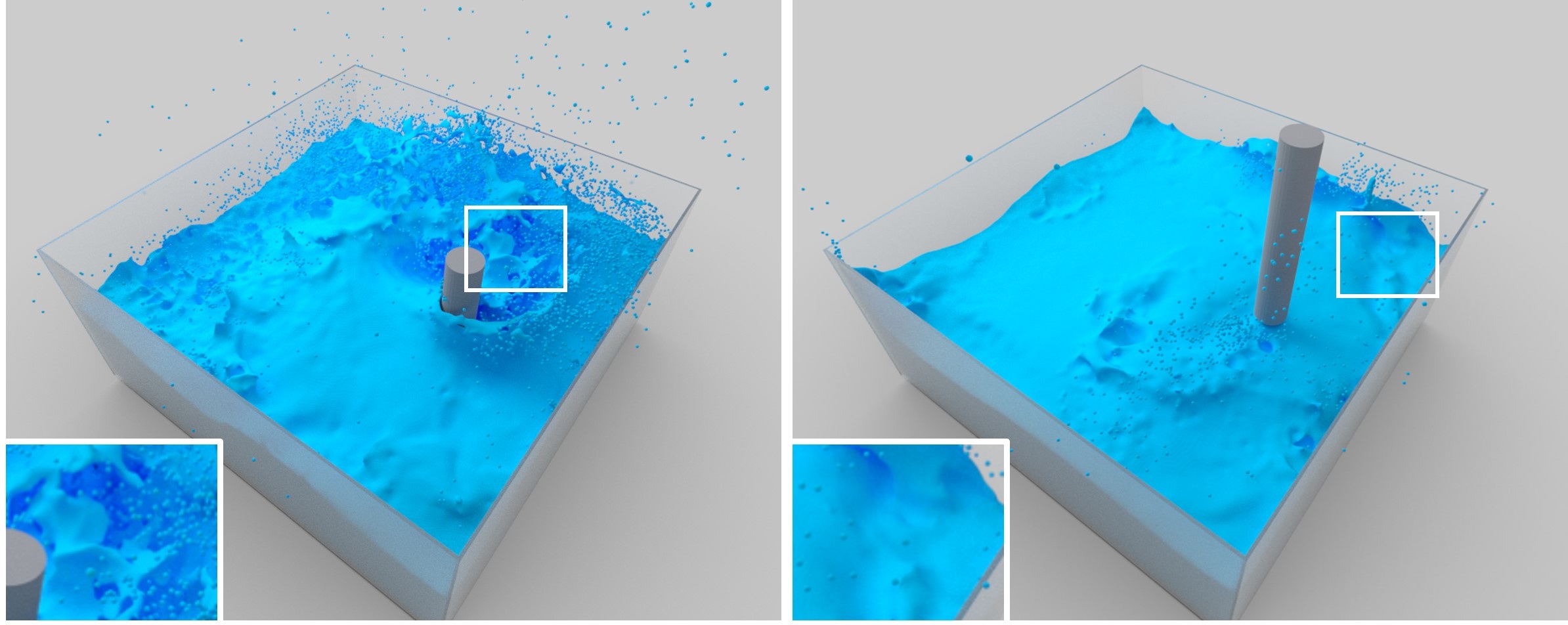}
\end{minipage}
}
\subfigure[Our method ($\alpha=1.2$)]{
\begin{minipage}[c]{1.0\linewidth}
\includegraphics[scale=0.207]{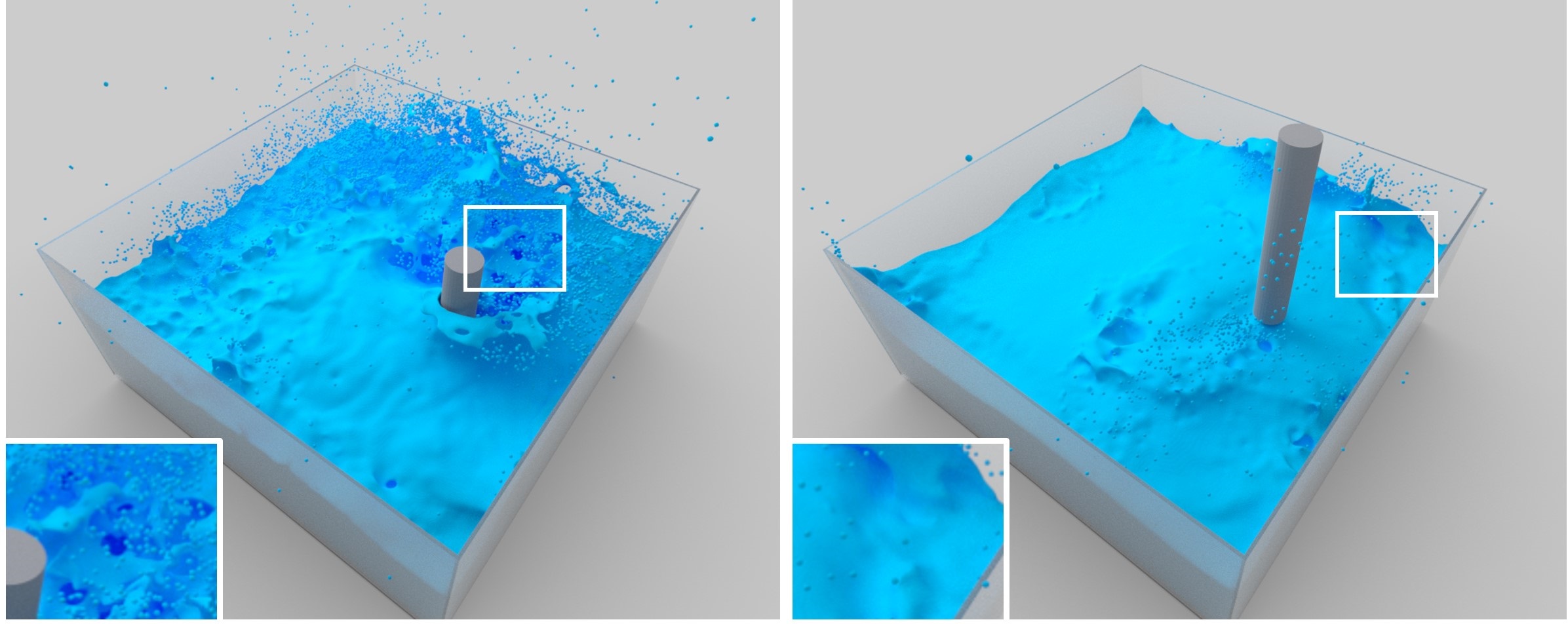}
\end{minipage}
}
 \caption{Comparison of DFSPH, the MP solver ($\nu_t=0.05$), and our method ($\alpha=1.0$ and $\alpha=1.2$) for a simulation of a stick (rod) mixing water. Besides turbulence enhancement, our method stands out in keeping the flow trail visible and maintaining the stability of the surface. As visible, the MP solver and our solver can obtain different enhancement results.} 
  \label{fig:stick} 
\end{figure}

Simulating incompressible fluids in DFSPH follows several steps, including advection and projection, and an extra divergence correction step which is applied to keep the velocity field divergence-free. The whole process is summarized in Algorithm~\ref{alg:one}, where $\Delta t$ denotes the size of one time step, ${\mathbf{a}_{adv} = \mathbf{a}_{vis} +\mathbf{a}_{g}}$, and $\mathbf{a}_{proj}$ and ${\mathbf{a}}_{correct}$ are the change rate of velocity derived form the implicit pressure field to satisfy the incompressibility and divergence-free conditions accordingly. Further, $\rho_{0}$ is the rest density of the fluid, and $\rho_{err}$, ${div}_{err}$, $n$, and $n'$ are user-specified scalar values as thresholds.


\begin{algorithm}[htb]
\noindent\textbf{Advectiom process:}\\
\par\setlength\parindent{2em}
\textbf{compute} $\mathbf{a}_{adv}$ \\
$\tilde{\bm{v}} := \bm{v}^{n} + \Delta t \mathbf a_{adv}$ \\
$\tilde{\bm{x}} := {\bm{x}^{n}} +  \Delta t {\tilde{\bm{v}}}$ \\
$\tilde{\rho} := positionBasedDensity(\tilde{\bm{x}})$ \\

\noindent\textbf{Projection process:}\\
while $(\tilde{\rho}-\rho_{0}) > \rho_{err} \ || \ numberOfIterations < n$\\
\qquad ${{p}} := positionBasedPressure({\tilde{\bm{x}}})$\\
\qquad $\mathbf{a}_{proj} := pressureBasedForce({p}) $\\
\qquad $\tilde{\bm{v}} := \tilde{\bm{v}} + \Delta t \mathbf a_{proj}$ \\
\qquad $\tilde{\rho} := positionBasedDensity(\tilde{\bm{x}} +  \Delta t {\tilde{\bm{v}}})$ \\
$\bm{x}^{n+1} = \tilde{\bm{x}}$ \\

\noindent\textbf{Divergence correction process:}\\
while $(\nabla\cdot \tilde{\bm{v}}) > {div}_{err} \ || \ numberOfIterations < n'$\\
\qquad ${p} := velocityBasedPressure({\tilde{\bm{v}}})$\\
\qquad $\mathbf{a}_{correct} := pressureBasedForce({p}) $\\
\qquad $\tilde{\bm{v}} := \tilde{\bm{v}} + \Delta t \mathbf a_{correct}$ \\
$\bm{v}^{n+1} := \tilde{\bm{v}}$ \\
\caption{Advection-projection with divergence correction}
\label{alg:one}
\end{algorithm}

\section{Vorticity Refinement Model for Turbulence Simulation}

Our method is closely related to Lagrangian vortex methods, namely it restores the velocity field through vorticity. In our method, besides velocity $\bm{v}$, each particle has a vector vorticity attribute $\bm{\zeta}$ defined as
\begin{equation}
\label{eqn:05}
\bm{\zeta} = \nabla \times \bm{v}.
\end{equation}
In a particle system, vorticity is a quantity used to describe the rotation of a particle. For the vorticity at the position of particle $i$, the value can be derived using Eqn.\@~\ref{diffSPH} as:
\begin{equation}
\label{eqn:06}
 \bm{\zeta}_{i} = \bm{\zeta}(\bm{v}_i) = \nabla \times \bm{v}_{i} = \sum_{j}\frac{m_{j}}{\rho_{j}}(\bm{v}_{i}-\bm{v}_{j})\times \nabla W_{ij}.
\end{equation}

%


\subsection{Vorticity Refinement}
Similarly to the divergence error issue \cite{bender2015divergence}, vorticity dissipation can also hinder the performance of a simulation. Recent SPH approaches~\cite{ihmsen2014implicit,bender2015divergence} for fluid animation can only correct negative divergence of the velocity field. As a result, the kinetic energy from the vorticity field is still allowed to be transformed into positive divergence during simulation, causing the loss of surface details and overall dynamic motions, effectively violating (the discrete version of) Eqn.~\ref{eqn:05}.

Given that the numerical dissipation of vorticity occurs between time steps, an ideal non-dissipative rate of change of vorticity is required to know the exact vorticity loss in each projection step. We achieve this through the curl of the Navier-Stokes equation (Eqn.\@\ref{eqn:01-1}) as:
%
\begin{equation}
\nabla \times \left(\frac{D\bm{v}}{Dt}\right) = {\frac{D\bm{\zeta}}{Dt}} = \bm{\zeta} \cdot \nabla \bm{v} + \nu_v\nabla^{2}\bm{\zeta},
\label{eqn:curl_ns}
\end{equation}
where $\bm{\zeta} \cdot \nabla \bm{v}$ is the stretching term, which is vital for physically meaningful turbulence motion evolution. We use Eqn.\@~\ref{eqn:curl_ns} to obtain the exact non-dissipative vorticity change of fluid particles between time steps, including boundary treatment~\cite{akinci2012versatile}. 

Note that $\bm{\zeta} \cdot \nabla \bm{v}$ in Eqn.\@~\ref{eqn:curl_ns} is a vector, which we compute, per coordinate, using the difference form of the SPH approximation (Eqn.\@~\ref{diffSPH}) via
\begin{equation}
\nabla v_i^{\{x,y,z\}} = \sum_j{
\frac{m_j}{\rho_j}
\left (
v_j^{\{x,y,z\}}-v_i^{\{x,y,z\}}
\right )
\nabla W_{ij}
},
\end{equation}

where $v_i^x$ is the $x$ component of the velocity of particle with index $i$, and similarly for $y$ and $z$. For the particle with index $i$, the vector $\bm{\zeta}_i \cdot \nabla \bm{v}_i$ can be thus derived as
\begin{equation}
    \bm{\zeta}_i \cdot \nabla \bm{v}_i
    =
    \left (
    \begin{array}{c}
    \bm{\zeta}_i \cdot \nabla v_i^x \\
    \bm{\zeta}_i \cdot \nabla v_i^y \\
    \bm{\zeta}_i \cdot \nabla v_i^z 
    \end{array}
    \right ).
\end{equation}
The Laplacian of $\bm{\zeta}$ in Eqn.\@~\ref{eqn:curl_ns} is derived using the artificial approximation analogous to Eqn.\@~\ref{eqn:04_1}. Hence, for the particle with index $i$, $\nu_v\nabla^{2}\bm{\zeta}_i$ can be derived as
\begin{equation}
\label{viscosity_cdot_laplacian_vectorZeta}
\nu_v\nabla^{2}\bm{\zeta}_i = 2(d+2)\nu_v \sum_{j} \frac{m_{j}}{\rho_{j}} \frac{\bm{\zeta}_{ij} \cdot \bm{x}_{ij} }{{{\bm{x}_{ij}}}^2 + 0.01 h^2} \nabla W_{ij}.
\end{equation}

%
%
%
According to Eqn.\@~\ref{eqn:curl_ns}, the ideal change of the vorticity field with respect to time, i.e., from time $t^n$ to $t^{n+1}$, is:
\begin{equation}
\label{idealVorticity}
\bm{\zeta}^{n+1} = \bm{\zeta}^{n} + \Delta t \frac{D\bm{\zeta}^{n}}{D t},
\end{equation}
%

%
%
and the dissipative vorticity update is given by:
\begin{equation}
\label{dissipatedVorticity}
\Delta\bm{\zeta} = \bm{\zeta}^{n+1} - \nabla \times \tilde{\bm{v}},
\end{equation}
where $\tilde{\bm{v}}$ is the (intermediate) velocity, as in the last line of Algorithm~\ref{alg:one}.


\begin{figure*}[htb]
\centering
\subfigcapskip=1pt 
\subfigbottomskip=2.5pt
\subfigtopskip=1pt
\subfigure[DFSPH]{
\begin{minipage}[c]{0.97\linewidth}
\includegraphics[scale=0.16]{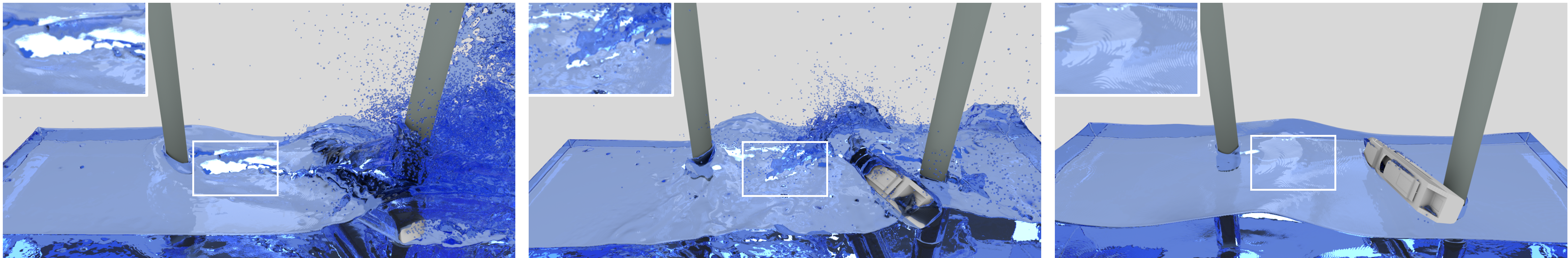}
\end{minipage}%
}

\subfigure[MP solver]{
\begin{minipage}[c]{0.97\linewidth}
\includegraphics[scale=0.16]{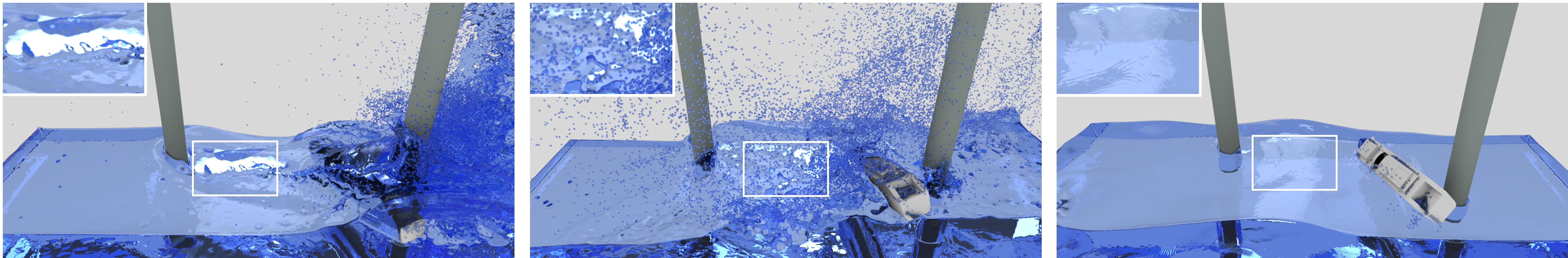}
\end{minipage}%
}

\subfigure[Our method]{
\begin{minipage}[c]{0.97\linewidth}
\includegraphics[scale=0.16]{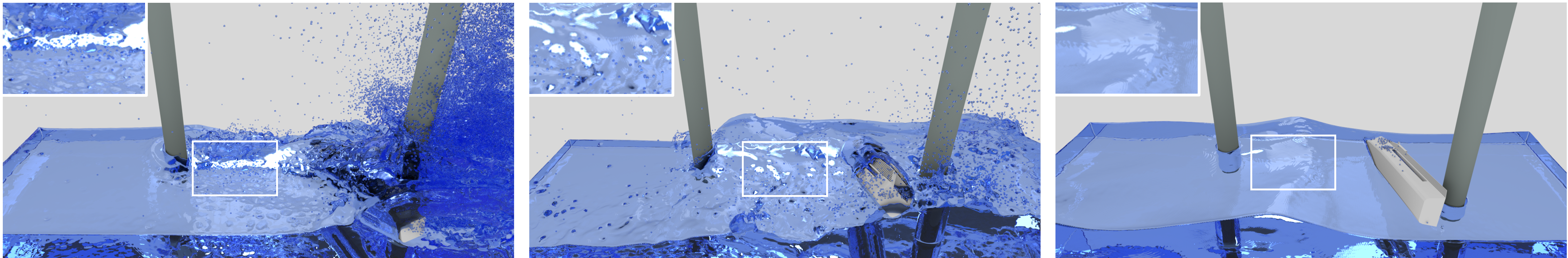}
\end{minipage}%
}
\caption{A breaking dam collides with a ship and 2 static pillars (1.70M particles). The key area is marked in white and zoomed in on. Our method and the MP solver have more details than the DFSPH solution. In the second column, the MP solver becomes unstable and some particles explode, while our method enhances the turbulence effect in a more stable and realistic way.
}
  \label{fig:boat} 
\end{figure*}

We next explain how we apply the update of Eqn.\@~\ref{dissipatedVorticity}. Assume that we know the velocity and position of all fluid particles at time $t^n$, and that the velocity at this time step is non-dissipative. We then get the velocity and position at time $t^{n+1}$ using the DFSPH approach. Next, we compute the vorticity at the current time $t^n$ and the next time $t^{n+1}$, denoted $\bm{\zeta}^{n}$ and $\bm{\tilde\zeta}$, respectively, from the velocity field using Eqn.\@~\ref{eqn:06}. By our assumption, $\bm{\zeta}^{n}$ is ideal, but $\bm{\tilde\zeta}$ is dissipative due to numerical integration. Thus the ideal vorticity value for a fluid particle at $t^{n+1}$, denoted $\bm{\zeta}^{n+1}$, is computed based on $\bm{\zeta}^{n}$ and the vorticity equation (Eqn.\@~\ref{eqn:curl_ns}). Hence, the dissipative vorticity value for this particle in Eqn.\@~\ref{dissipatedVorticity} can be converted to $\Delta \bm{\zeta} = \bm{\zeta}^{n+1} - \bm{\tilde\zeta}$. The dissipated vorticity is used to refine the velocity using the stream function, as explained next.


\subsection{Solving Velocity via the Stream Function}



Inspired by~\cite{zhang2014pppm}, we express the relationship between the velocity $\bm{v}$ and the vorticity $\bm{\zeta}$ using the stream function $\bm{\psi}$ as:
\begin{equation}
\label{streams}
\begin{aligned}
\bm{v} =& \nabla \times \bm{\psi}, \\
\nabla^{2}\bm{\psi} &= -\bm{\zeta}.
\end{aligned}
\end{equation}
Green’s function provides a semi-analytical solution for the stream function. The derivation from the stream function to linear velocity can be solved using Eqn.\@~\ref{diffSPH}. Generalized by the Helmholtz decomposition, the stream function is the vector potential $\bm{\psi}$ of the velocity field $\bm{v}$, which can be defined as
\begin{equation}
\label{continusStream}
   \bm{\psi}(\bm{x}) = \int_{\mathbb{R}^{3}} {\frac{\nabla \times \bm{v}(\bm{y})}
{4\pi \left \| \bm{x}-\bm{y} \right \|}},
\end{equation}
that is, the stream function $\bm{\psi}$ at position $\bm{x}$ is computed by integrating the curl of velocity $\bm{v}$ at position $\bm{y}$ over the three-dimensional space $\mathbb{R}^3$. 
Using Eqn.\@~\ref{eqn:05}, we next discretize Eqn.\@~\ref{continusStream} to get the stream function at the local position of particle with index $i$ as:
\begin{equation}
\label{eqn:discreteStream}
    \bm{\psi}_{i} = \frac{1}{4\pi} \sum _{\| \mathbf{x}_i - \mathbf{x}_j \| \leq h} \frac{\Delta \bm{\zeta}_{j} {}{V}_{j}}
{\left \| \bm{x}_{i}-\bm{x}_{j} \right \|},
\end{equation}
where $V_j$ stands for the volume represented by the particle with index $j$.
Ideally, $V_j$ should be infinitely small and all distances $\|\bm{x}_{i}-\bm{x}_{j}\|$ should be considered in the summation in Eqn.~\ref{eqn:discreteStream}. However, to limit computational overhead and its adaptability to SPH, we only include neighbouring particles within a smoothing radius $h$ in Eqn.\@~\ref{eqn:discreteStream}. This is justified by the fact that the influence of neighbour particles shrinks with distance. Although the approximation could potentially induce instability and dissipation, our results show that this improves performance without sacrificing turbulent details, as already observed e.g.\ in~\cite{muller2003particle}.

With the stream function obtained for each particle, the refined velocity for the particle with index $i$ is derived as 
\begin{equation}
\Delta\bm{v}_i=\sum_j{
\frac{m_j}{\rho_j}
\left (
\bm{\psi}_i-\bm{\psi}_j
\right )
\times \nabla W_{ij}}.
\end{equation}



To extend the flexibility of our method, we introduce an adjustment parameter \cbl{$\alpha \in R$}, with the default value of 1 representing the ideal vorticity refinement. It controls the amount of turbulence added to every simulation time step. Therefore the refined linear velocity at $t^{n+1}$ is expressed as

\begin{equation}
\label{refined_with_alpha}
    \bm{v}^{n+1}=\tilde{\bm{v}}+\alpha\Delta\bm{v}.
\end{equation}

Since the divergence of the curl of any field is zero, the correction of linear velocity due to vorticity does not cause any further divergence deviations. Hence, our method does not contradict any SPH principles, making it easier to implement into standard Lagrangian approaches. Algorithm~\ref{alg:two} summarizes our method, integrated with the DFSPH technique for SPH simulation; see also Fig.\@~\ref{fig:my_label}.

\begin{algorithm}[!h]
\SetAlgoNoLine

\noindent\textbf{Compute current vorticity field:} 
\ \ \ $\bm{\zeta}^n = \nabla\times\bm{v}^n$ \\



\noindent\textbf{Advection-projection:} 
\qquad \qquad \ \ \ \ $\overline{}{\bm{v}}^{adv} = advectProject(\bm{v}^n)$ \\

\noindent\textbf{Correct divergence field:} 
\qquad \qquad $\tilde{\bm{v}} = correctDivergence(\overline{}{\bm{v}}^{adv})$ \\

\noindent\textbf{Vorticity through linear field:} 
\ \ \ \ \quad $\tilde{\bm{\zeta}} = \nabla \times \tilde{\bm{v}}$ \\

\noindent\textbf{Compute vorticity equation:} 
\ \ \ \ \ \ \quad $\bm{\zeta}^{n+1} = \bm{\zeta}^{n} + \Delta t \frac{D\bm{\zeta}^{n}}{D t}$ \hfill (Eqn.\@~\ref{idealVorticity}) \\

\noindent\textbf{Dissipation of vorticity:} 
\qquad \qquad \ \ $\nabla \times \bm{v}(\bm{y}) = \bm{\zeta}^{n+1} - \tilde{\bm{\zeta}}$ \hfill (Eqn\@.~\ref{dissipatedVorticity})\\

\noindent\textbf{Compute stream function:} 
\ \ \ \ \ \ \ \ \ \ \ \ \   $\bm{\psi} = \int_{\mathbb{R}^{3}} {\frac{\nabla \times \bm{v}(\bm{y})}
{4\pi \left \| \bm{x}-\bm{y} \right \|}}$ \hfill
(Eqn.\@~\ref{continusStream}) \\

\noindent\textbf{Refinement of linear velocity:} 
\quad \ \ \ \ $\Delta\bm{v} = \nabla \times\bm{\psi} $ \hfill (Eqn.\@~\ref{streams})\\

\noindent\textbf{Refine linear velocity:}
\qquad \qquad \ \ \ \ \ $ \bm{v}^{n+1} = \tilde{\bm{v}} + \alpha \ \Delta\bm{v} $ 
\hfill (Eqn.\@~\ref{refined_with_alpha})\\

\caption{Our vorticity refinement (VR) solver}
\label{alg:two}
\end{algorithm}


\section{Results and Discussion}
We next test our novel Vortex Refinement (VR) method on several scenes, comparing it with the state-of-the-art micropolar (MP) model and classical SPH approaches.

Both the VR and the MP method are integrated with DFSPH in the following experiments to show the applicability of our method. We used the boundary handling method proposed by Akinci et al. \cite{akinci2012versatile}. We implemented the entire framework in C++, with animations rendered by Blender. Our simulation platform is a graphic workstation with an Intel Xeon E5-2687w v4 (15M cache, 3.5 GHz, 12 cores) CPU, 80 GB RAM, and an NVIDIA Quadro P4000 GPU.

\begin{figure*}[htb]
\centering
  \includegraphics[scale=0.7]{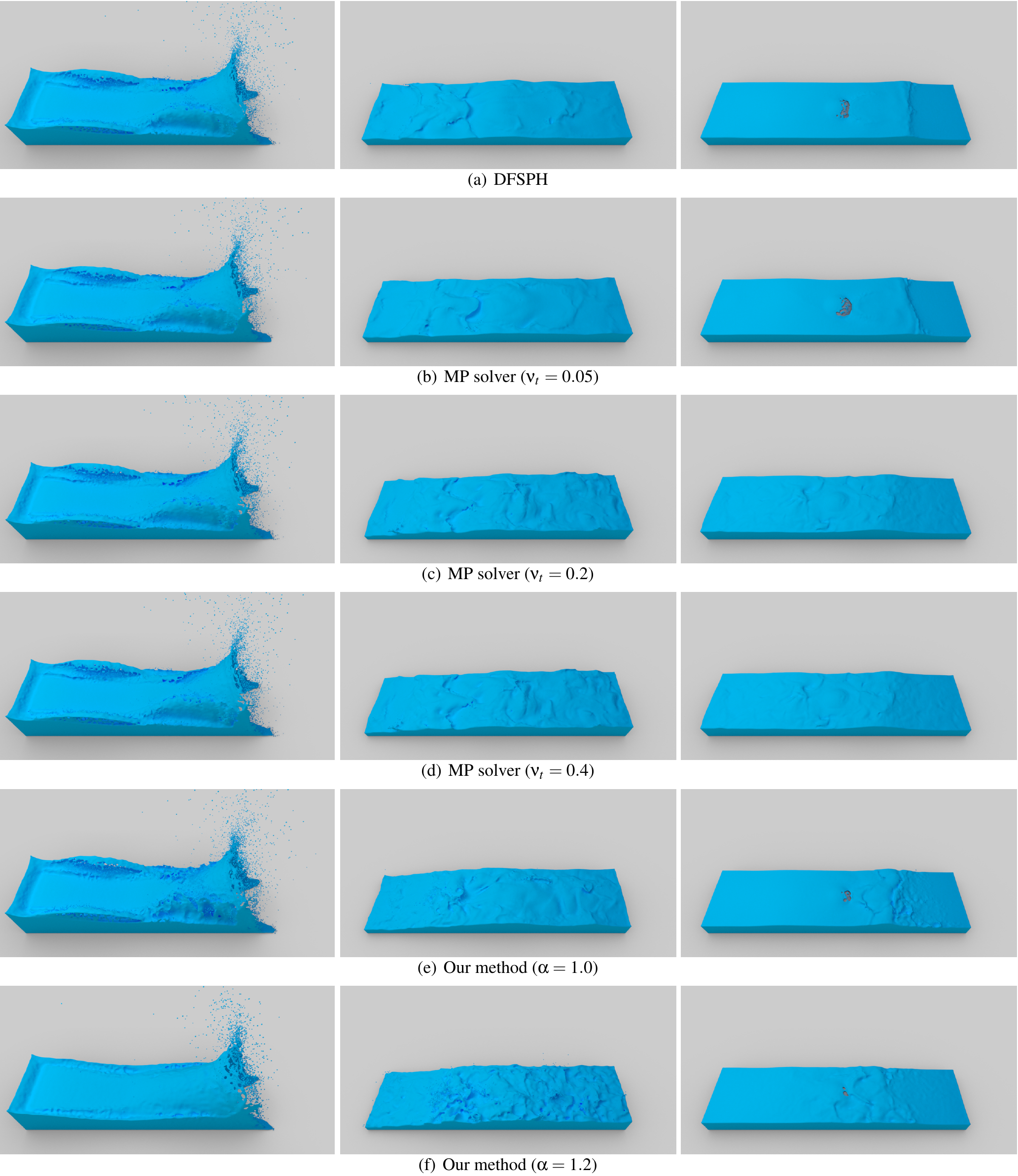}
  \caption{Comparison of DFSPH, the MP solver with ${\nu_t= 0.05, 0.2, 0.3, 0.4}$, and our method with $\alpha = 1.0, 1.2$ in the breaking dam scenario with a static spherical obstacle placed to the right.}

  \label{fig:ball_new} 
\end{figure*}

Similarly to the adjustment parameter $\alpha$ in our method, there is a scalar $\nu_t$ in the MP method to control it. Based on the mechanism of the MP method~\cite{eringen1966theory,bender2018turbulent}, $\nu_t$ greater than $\nu_v$ can potentially violate the second law of thermodynamics. In all experiments below we set $\nu_v=0.05$. We therefore choose $\nu_t=0.05$ as a natural refinement for the MP solver, which corresponds to $\alpha=1$ in our method. However, to explore the stability and performance of the methods, we test $\alpha$ greater than $1$ and $\nu_t$ greater than $0.05$; see Figs.\@~\ref{fig:stick} and~\ref{fig:ball_new}. As stated in ~\cite{bender2018turbulent}, fluids are reasonably stable when $\nu_t \le 0.4$. 

\begin{figure*}[tbp]
\centering
\begin{tabular}{c@{\,\,}c}
\includegraphics[scale=0.33]{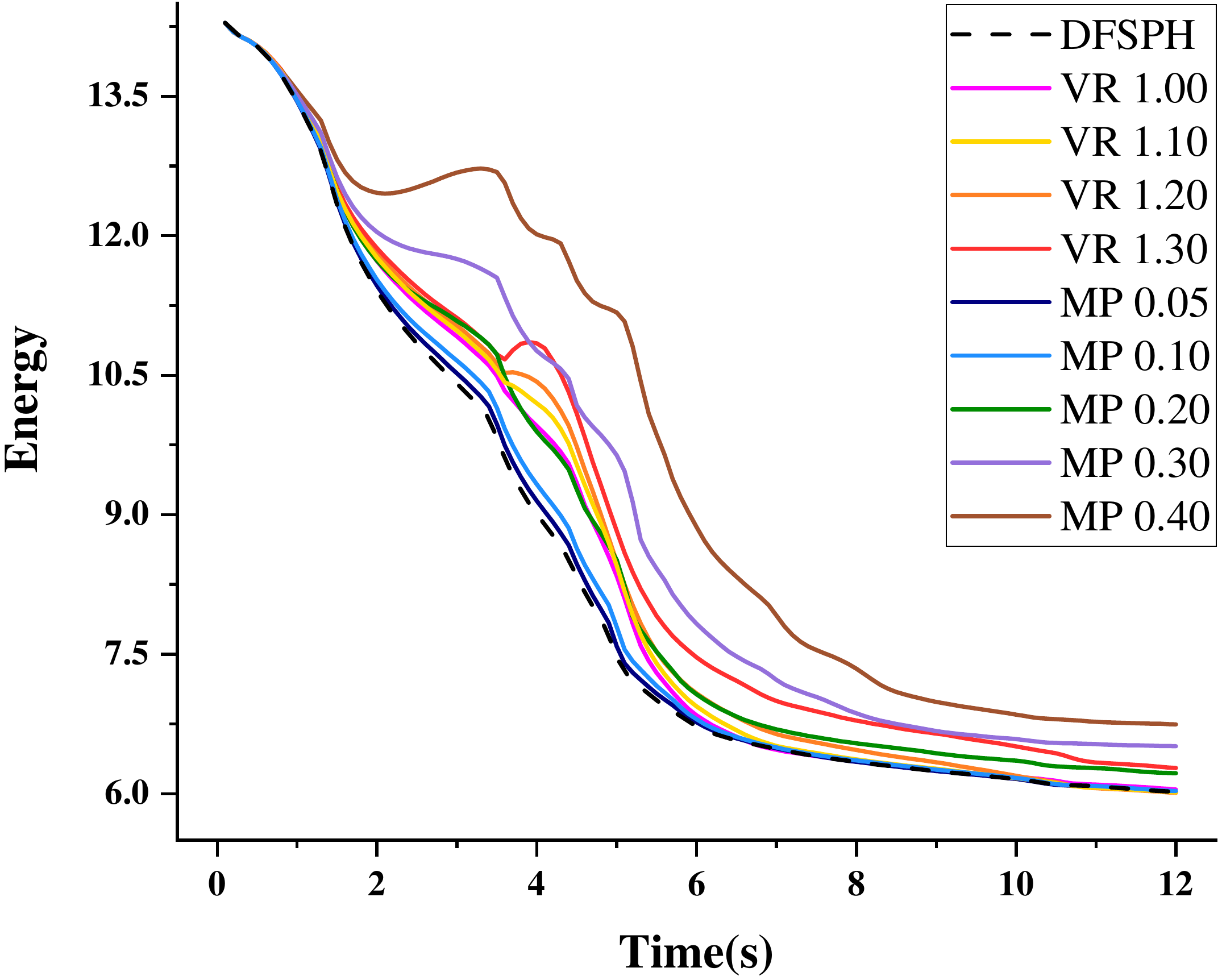} &
\includegraphics[scale=0.33]{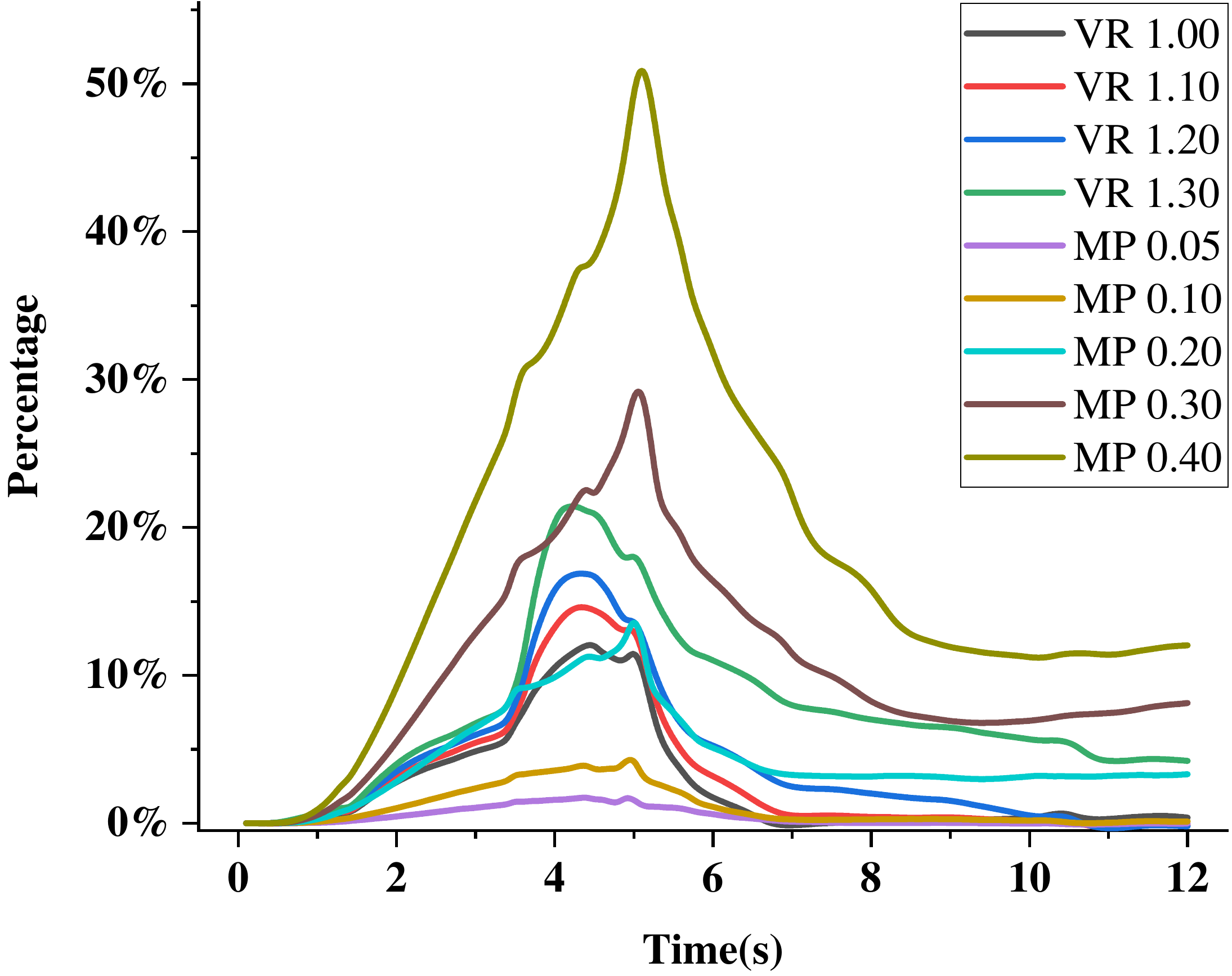}
\end{tabular}
\caption{Comparison of energy changes for different methods using different parameter values in the breaking dam scene shown in Fig.\@~\ref{fig.breakdam.energy}. Left: direct energy comparison. Right: energy increase ratio relative to DFSPH. When the fluid is flowing ($t\in [2,8]$), our method and the MP solver are able to add energy to the scene and enhance the visual effect. However, the MP solver with $\nu_t = 0.2, 0.3, 0.4$ and our method with $\alpha = 1.3$ do not converge after $t = 8$ due to the excessive energy added.}
    \label{fig:energy}
\end{figure*}

\begin{figure}[tbp]
\centering
\begin{tabular}{c@{\,}c}
\includegraphics[scale=0.14]{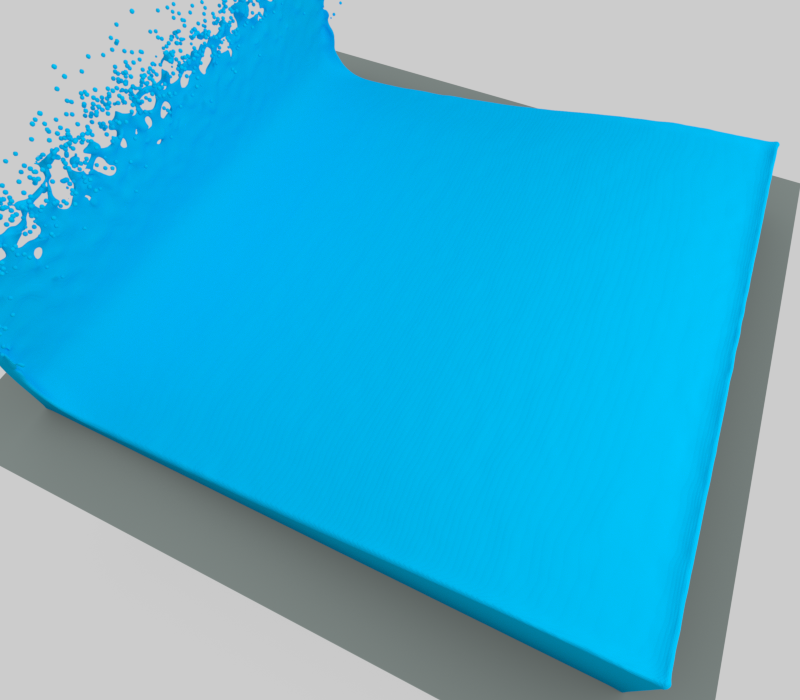} &
\includegraphics[scale=0.14]{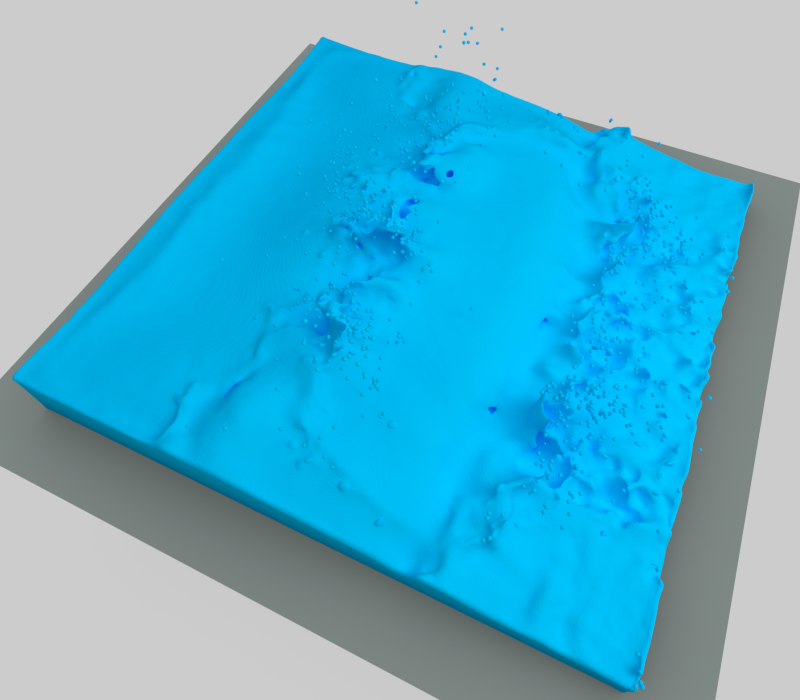}
\end{tabular}
\caption{A breaking dam scenario for the evaluation of energy changes (461K fluid particles); see Fig.\@~\ref{fig:energy}.}
\label{fig.breakdam.energy}
\end{figure}



\subsection{Effectiveness and Comparison}
To show the effectiveness of our approach numerically, we executed two breaking-dam experiments, and we executed two other experiments for parameter discussion and energy comparison with other methods, as follows. 

\noindent\textbf{Breaking dam with a board.}
In Figs.\@~\ref{fig:abs} and \ref{fig:board}, a board collides with a breaking dam which only allows fluid to go through the so-created gap. Figure~\ref{fig:board} shows the results with 1.18M particles. Only few vortex effects can be seen using DFSPH. Water flushes through the gap and dissipates quickly without clear turbulence effects. Compared to DFSPH, our solver generates several realistic vortices around the board and corners. The MP solver also improves the visual result, but not as obviously as our method. Since our method refines particle velocity based on the vorticity field, vortices are naturally preserved and turbulence is generated from the dissipated energy in a realistic way. In Fig.\@~\ref{fig.particle vorticity}, the vorticity magnitude of all particles is visualized. The comparison shows that both our method and the MP method yield  higher energy values than DFSPH. The MP solver adds energy in a natural way, while our method  recovers energy from numerical dissipation more effectively and is thus able to simulate more details.
\begin{figure}[h]
    \centering
    \includegraphics[scale=0.35]{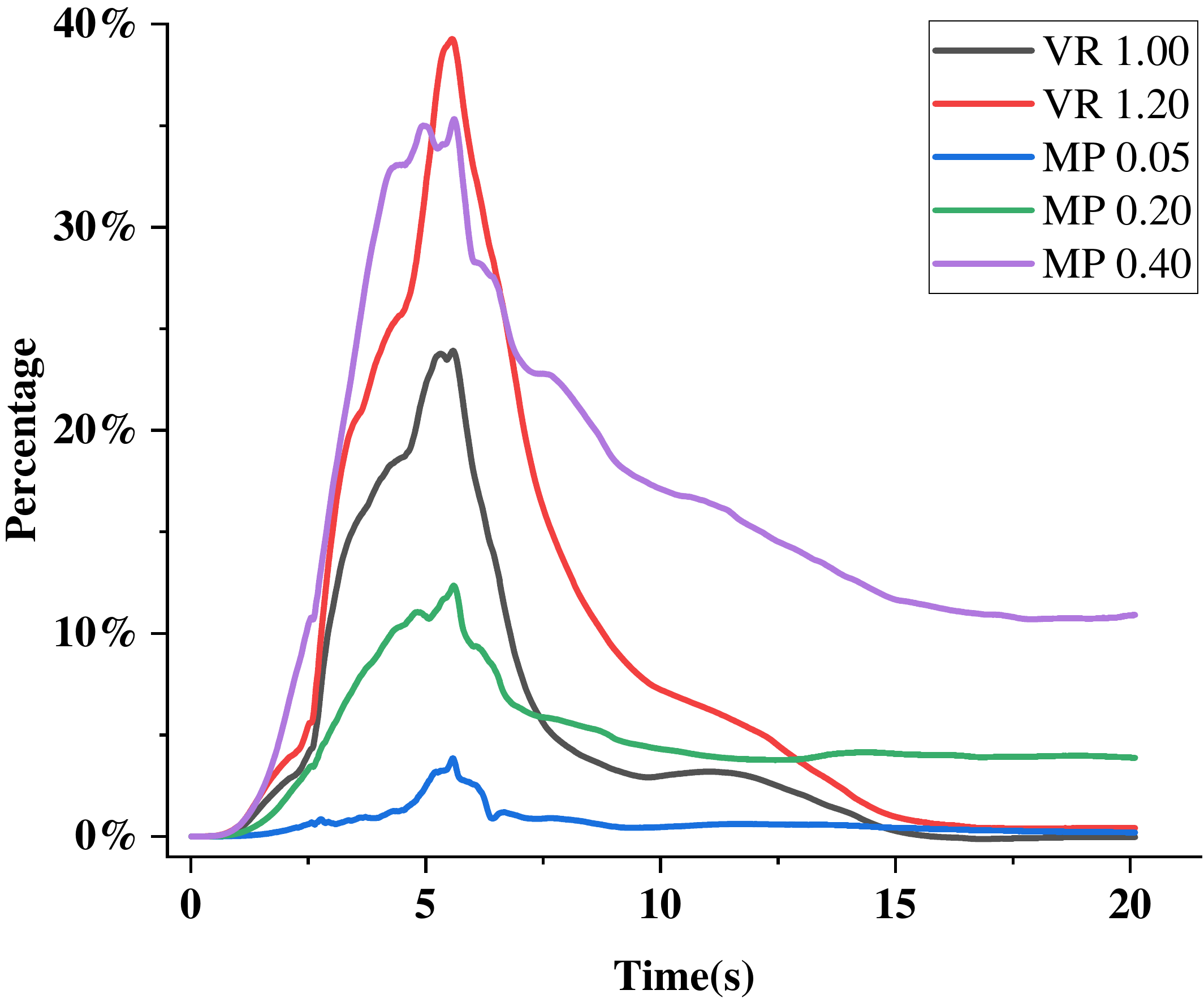}
    \caption{Similar to Fig.\@~\ref{fig.breakdam.energy}, we compared the energy increase ratio of different methods to the DFSPH method using the scene shown in Fig.\@~\ref{fig:ball_new}. We see that all methods produce more energy than DFSPH when the fluid is flowing ($t \in [0, 15]$). Note that the MP solver with $\nu_t = 0.2, 0.4$ does not converge after $t = 15$.}
    \label{fig:cqenergy}
\end{figure}

\begin{figure}[tbp]
\centering  
\subfigure[DFSPH]{
\begin{minipage}[c]{0.5\linewidth}
\includegraphics[scale=0.66]{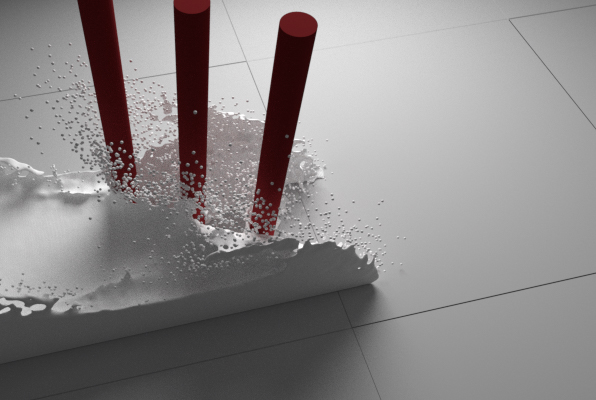}\vspace{4pt}
\end{minipage}%
\begin{minipage}[c]{0.5\linewidth}
\includegraphics[scale=0.66]{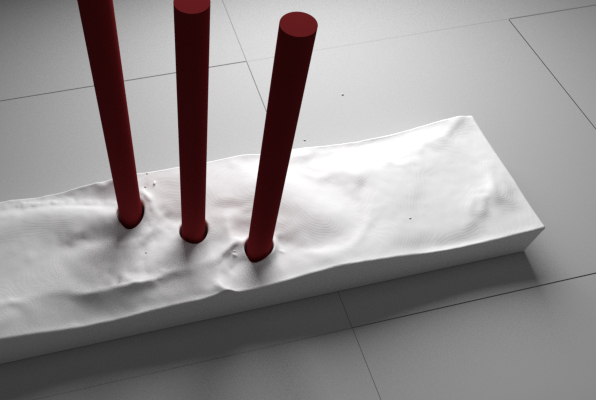}\vspace{4pt}\end{minipage}%
}\\[-1mm]
\subfigure[MP solver]{
\begin{minipage}[c]{0.5\linewidth}
\includegraphics[scale=0.66]{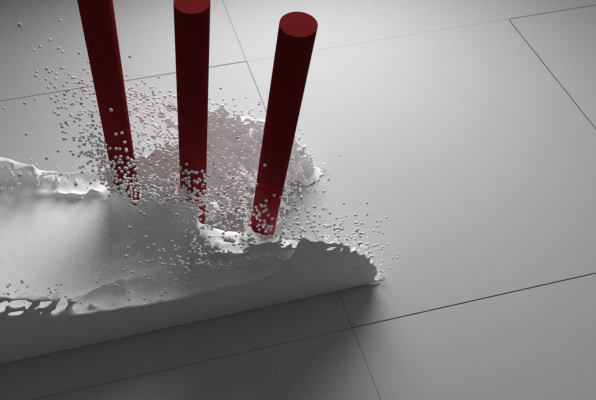}\vspace{4pt}
\end{minipage}%
\begin{minipage}[c]{0.5\linewidth}
\includegraphics[scale=0.66]{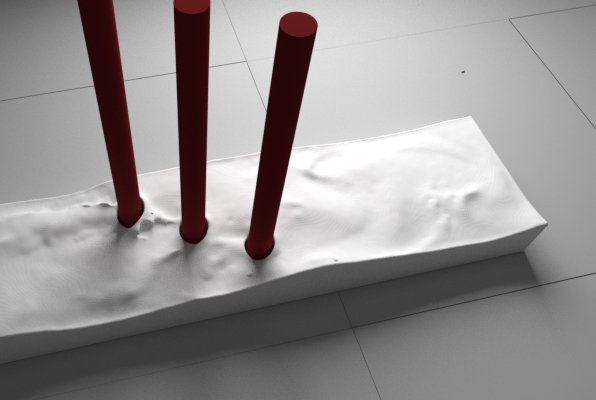}\vspace{4pt}\end{minipage}%
}\\[-1mm]
\subfigure[Our method]{
\begin{minipage}[c]{0.5\linewidth}
\includegraphics[scale=0.66]{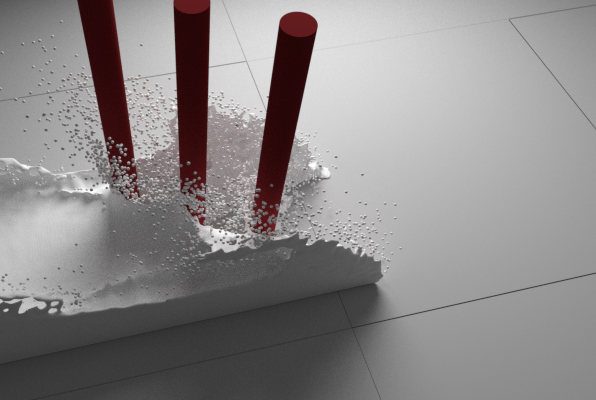}\vspace{4pt}
\end{minipage}%
\begin{minipage}[c]{0.5\linewidth}
\includegraphics[scale=0.66]{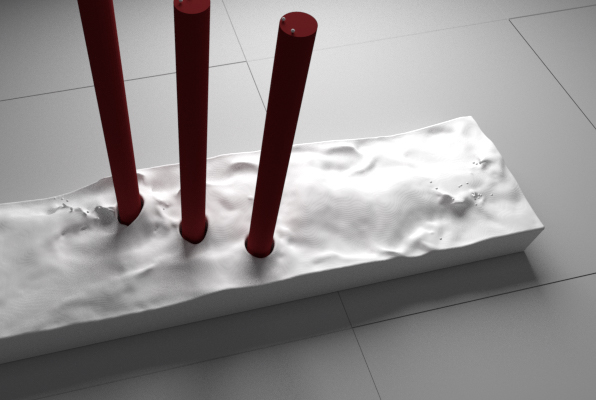}\vspace{4pt}\end{minipage}%
}
\caption{A comparison of DFSPH, the MP solver and our method in a breaking dam scene with 3 static pillars as obstacles. The water hits the cylinders and the right wall and bounces back, forming turbulence in the process. Compared to DFSPH, our solver gives rise to tiny vortices instead of the water simply going around the pillars. Compared to the MP solver, our method seems to provide more turbulent details but smaller vortices in this scene.}
  \label{fig:obstacle} 
\end{figure}

\noindent\textbf{Breaking dam with three obstacles.}
As shown in Fig.~\ref{fig:obstacle}, a breaking dam scenario with static obstacles was tested using 457K fluid particles. The fluid flows in from the left and hits the wall on the right. Several waves are generated in the process, which then come back and interact with three rigid bodies. Desirable turbulence can be observed over the surface. We compared our method with the DFSPH and MP solvers. In DFSPH, the fluid seems to go around the pillars and forms splashes, but scarcely any complex turbulence effects. In contrast, our solver creates small-scale vortices instead of just the fluid smoothly flowing around the pillars. Since these small vortices cannot sustain a self-spinning state, they quickly break down into turbulence. Compared to the MP solver in this scene, our method seems to generate more turbulent details but smaller vortices. The MP solver and our method can achieve different visual effects.

\noindent\textbf{Energy Comparison.} An energy comparison of a breaking dam experiment (see Fig.\@~\ref{fig.breakdam.energy}) is shown in Fig.\@~\ref{fig:energy}. The left plot shows the energy comparison, while the right plot shows the energy increase ratio relative to DFSPH. When $t \in [2,8]$, the fluid keeps flowing and forming turbulence. If the energy is larger than that of DFSPH, then energy is recovered (or added) successfully. After the water surface calms down (after about 10s), the scene should contain only potential energy (no kinetic energy). The energy of the traditional DFPSH method can be used as a benchmark: If a method generates, at this time point, more energy than DFSPH, then this method is considered to create \emph{additional} energy. In this comparison experiment, the energy values after 10s for both the MP solver with $\nu_t = 0.05, 0.1$ and for our method with $\alpha = 1.0, 1.1, 1.2$ are very close to the DFSPH values. Our method with $\alpha = 1.3$ and the MP method with $\nu_t = 0.2, 0.3, 0.4$ have higher energy than DFSPH. 
In some applications, in order to enhance the visual effect, one can use such larger parameter values. However, this can very likely cause excessive chaos and even instability such as unnatural turbulence similar to boiling. Hence, we recommend to use our method with $\alpha = 1.0$ to ensure the energy is always in line with the underlying physics.

\noindent\textbf{Breaking dam with a hemisphere: parameter influence.}
In this experiment (see Fig.\@~\ref{fig:ball_new}) we flush a hemisphere obstacle with a fixed volume of fluid. This means only limited kinetic energy is involved in this scenario (from gravitational potential energy). We simulated the flow using DFSPH, our method with $\alpha=1$ and $\alpha=1.2$, and the MP solver with $\nu_t=0.05$, $\nu_t=0.2$ and $\nu_t=0.4$. 
When comparing the DFSPH approach with our method with $\alpha=1$ and with MP with $\nu_t=0.05$, both methods are able to increase the turbulence performance, but our result is more pronounced than the MP one. To obtain more obvious turbulence effects, we increase the turbulence control parameters in the two methods, which means that more energy is added to the simulation. The renderings show that our method with $\alpha = 1.2$ yields more turbulence and the result is better than that of the MP solver with $v_t = 0.2$. To keep our method in line with the underlying physics, as explained for the earlier example, we do not use higher parameter values.
The MP solver adds more turbulence in this scene. The obtained results are visually more salient for large  parameter values, \emph{e.g.} $\nu_t=0.2$. However, $\nu_t$ cannot be increased indefinitely. For example, if we set $\nu_t=0.4$ (Fig.\@~\ref{fig:ball_new}, last row), the fluid does not calm down, which is unnatural. The detailed energy comparison is shown in Fig.\@~\ref{fig:cqenergy}. Our method can be applied to scenes that are more sensitive to physics laws, such as adding more details to a relatively stably-flowing scene. In contrast, the MP method can be used in scenes where one wants to create a stronger visual impact, such as collapses or violent shocks.

Overall, this experiment shows that the MP solver and our solver can achieve different turbulence effects. Our method achieves better turbulence results without adding energy sources. In contrast, the MP solver can add small vortices, but when increasing its parameter values, energy sources will pop up and prevent the fluid from calming down.

\begin{figure*}[htb]
\centering  
\centering
\subfigcapskip=2pt 
\subfigbottomskip=0pt
\subfigtopskip=0pt
\subfigure[DFSPH]{
\begin{minipage}[c]{0.97\linewidth}
\includegraphics[scale=0.235]{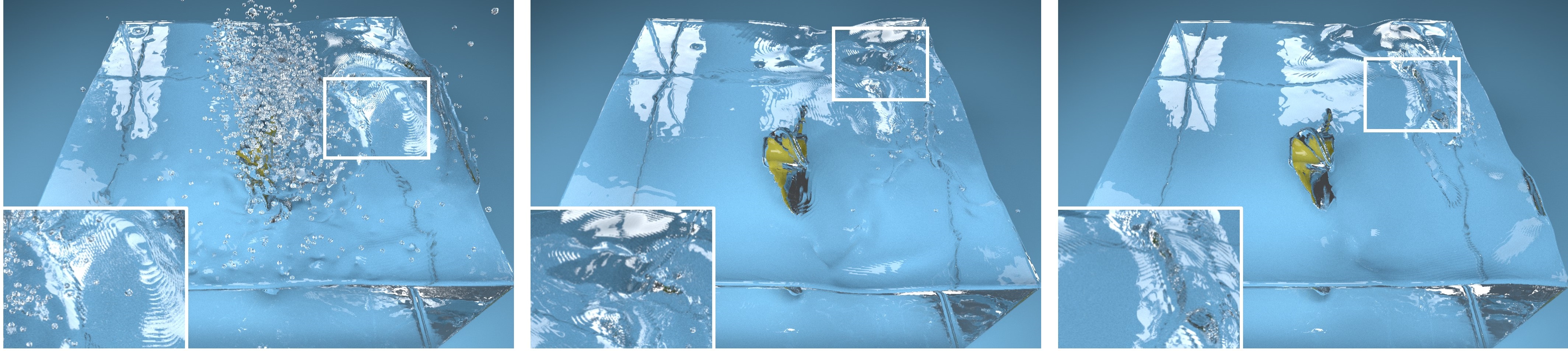}
\end{minipage}%
}
\subfigure[MP solver]{
\begin{minipage}[c]{0.97\linewidth}
\includegraphics[scale=0.235]{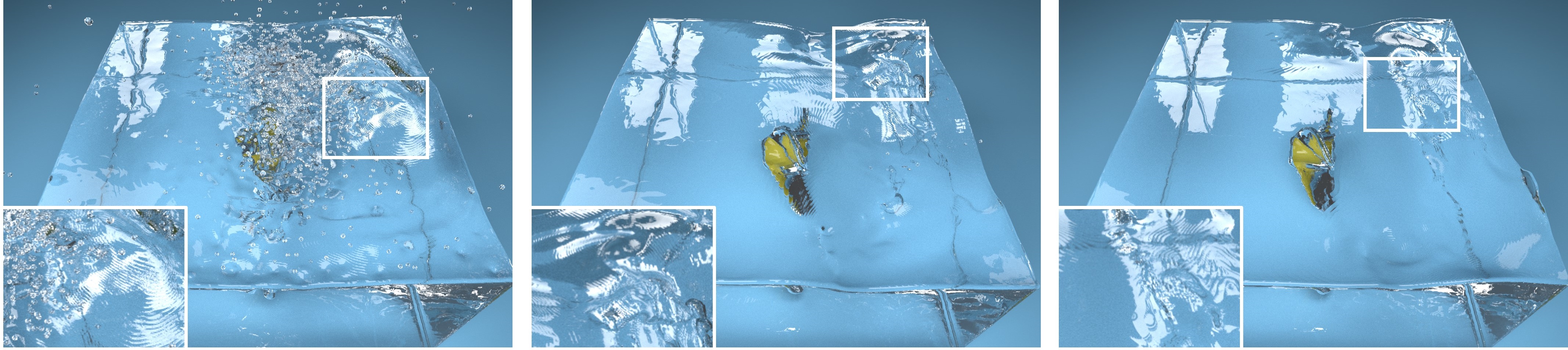}
\end{minipage}%
}
\subfigure[Our method]{
\begin{minipage}[c]{0.97\linewidth}
\includegraphics[scale=0.235]{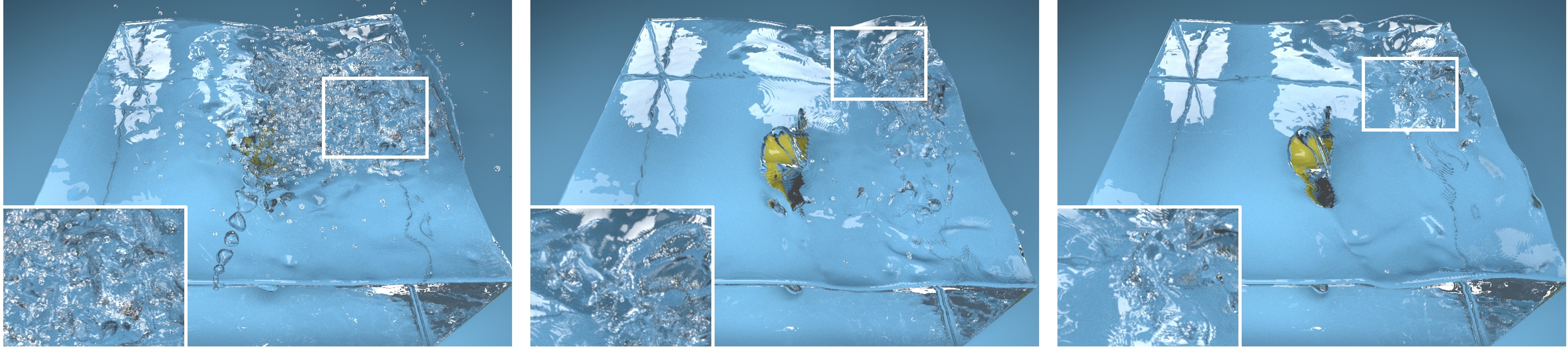}
\end{minipage}%
}

\caption{A propeller interacts with 1.29M fluid particles using DFSPH, the MP solver, and our method. DFSPH (top row) is not able to produce a complex flow. Under stable conditions, the MP solver (middle row) can only add limited turbulence effects to DFSPH, while our VR method is able to generate realistic vortices (bottom row). The improved turbulence performance can be seen clearly in the insets. 
}

  \label{fig:propeller} 
\end{figure*}

\subsection{Quality}
%

To further demonstrate the turbulence quality of our method, we simulated several complex scenarios with dynamic boundary conditions and compared them with the MP solver. 

\noindent\textbf{Spinning Propeller.} A propeller is slowly submerged into water, after which it starts spinning at 3 radians per second. Fig.\@~\ref{fig:propeller} shows the results of this simulation using 1.29M fluid particles for DFSPH, MP, and VR (our method).
Observe that neither the complex flow nor strong turbulence effects are produced and preserved using DFSPH. Both our method and the MP method enhance the visual effect. In contrast to the MP method, our method adds energy in a physically reasonable way (no turbulence in front of the propeller) and creates vivid turbulent details over the free fluid surface. The key areas are zoomed in on. Also, a vortex is observed with our method after the propeller has stopped spinning (see also the supplementary video).

\noindent\textbf{Boat-sinking.} In this scenario, a boat and two columns interact with a breaking dam. Figure~\ref{fig:boat} shows the results using 1.7M fluid particles. The potential energy of the fluid transforms into the kinetic energy of the fluid particles and the boat. The water is first violently displaced when it hits the column and the boat, and next gradually calms down as time goes by, finally reaching a stable state. We see that the DFSPH method produces relatively weakly turbulent details, which get lost quickly due to numerical dissipation. In contrast, our method and MP server shows more natural dynamics with realistic turbulent effects on the fluid surface. The fluid gradually calms down as time goes on. Our method and the MP method achieve different styles.

\noindent\textbf{Stirring water.} In Fig.~\ref{fig:stick}, a cylindrical stick was inserted into a tank of water, and stirred at a uniform speed for several seconds. The water splashed around due to the quick movement of the stick. Observe that the trace left on the surface lasts longer in our method than with the MP method, which is a critical point for boat-sailing animation scenarios. After the stirring process, the stick is pulled out of the fluid, and the water starts to calm down. The DFSPH approach calms the fluid down quickly due to numerical dissipation. The surface details are clearer and sharper in our method. Also, we notice a disturbance wave in the MP method, caused by the fact that $\nu_t$ exceeds the kinematic viscosity.



The above three scenarios show that our method can keep stability when dealing with extreme conditions like strong collisions, while physically preserving energy. Moreover, in the accompanying video it can be seen that our method not only amplifies existing vortices but also generates new ones.

\noindent\textbf{Computational overhead.} The computational overhead of our method is negligible compared to the whole SPH simulation procedure. Table~\ref{tab:time} shows the computing times for DFSPH, the MP solver, and our method for different simulation scenes. The different computation times are explained as follows.
Compared to DFSPH, both turbulence methods (MP and ours) need to compute the vorticity field, i.e., solve for the Laplacian $\nabla^2\bm{\zeta}$. Further, $\nabla\times\bm{\zeta}$ (in the MP solver) and $\nabla\bm{v}$ (in our method) also need to be solved for. The difference is that our method needs to compute $\bm{\psi}(\bm{\zeta})$ and $\nabla\times\bm{\psi}$ to get the refined velocity, but as Table~\ref{tab:time} shows, the extra computational effort is negligible.


\begin{table}[]
    \centering\small
    \begin{tabular}{|@{\,}l@{\,}|@{\,}r@{\,}||@{\,}r@{\,}|@{\,}r@{\,}|@{\,}r@{\,}|@{\,}r@{\,}|@{\,}r@{\,}|@{\,}r@{\,}|}
    \hline
        Experiment & Fig. & Particles & $\Delta t$ (ms) & Steps & DFSPH (m) & MP (m) & VR (m)\\
        \hline
        \hline
         Board & \ref{fig:board} & 1.18M & 2.4 & 9542 & 2401.2 & 2565.9 & 2565.1 \\
        \hline
         Stirring & \ref{fig:stick} & 1.39M & 2.4 & 9542 & 2399.9 & 2864.4 & 2693.4 \\
        \hline
         Sphere & \ref{fig:ball_new} & 899.8K & 2.4 & 8375 & 1657.4 & 1715.2 & 1827.2 \\
        \hline
         Pillars & \ref{fig:obstacle} & 457K & 3 & 6667 & 156.4 & 188.3 & 218.5 \\
        \hline
         Propeller & \ref{fig:propeller} & 1.29M & 3 & 7334 & 1782.1 & 2309.1 & 2338.9 \\
        \hline  
    \end{tabular}
     \caption{Total time comparisons of three methods: DFSPH, MP, and our method (VR) over five simulations. $\Delta t$, in milliseconds, is the time step used in the experiments, and the total computation times, in minutes, include the costs of the density solver and the divergence-free solver in DFSPH.}
    \label{tab:time}
\end{table}

\section{Conclusion and Discussion}
We have presented a particle-based turbulence refinement method that recovers lost velocity from the difference between the theoretical and the actual vorticity value. Our method can not only increase existing vortices significantly by recovering numerical dissipation, but also generates new turbulence at potentially different locations. The turbulence-enhancement parameter of our method has a theoretically optimal value ${\alpha = 1}$ that can increase turbulence without adding too much energy. At the same time, one can easily adjust this parameter to achieve different turbulence levels for different simulation effects.

Experimental results show that, compared to the classical and micropolar SPH methods, our method is able to enhance turbulent effects more visibly. Furthermore, our method guarantees energy conservation, even when using a large particle radius and/or a large time step. This means that our method is still robust even under extreme simulation conditions and can handle complex large-scale scenes, as demonstrated in our simulation scenarios.

Numerical dissipation is difficult to fully correct in SPH methods. Our method can simulate typical turbulent scenes efficiently and is relatively stable even for scenarios with highly turbulent flow. At the same time, we should note that some vorticity is lost in such cases. While this small amount of loss does not affect the general visual quality, decreasing it is an open topic for future research, which can be expected to lead to even more realistic fluid simulations.

In the future, we aim to investigate merging our method with microstructural models, since these models show great potential for rough simulation conditions and also have a close relationship with viscosity. Improving computation accuracy is another potential future research direction. Finally, increasing the computational scalability of our method by e.g.\ efficient and effective parallelization is attractive for making our method directly applicable to complex real-world and/or interactive simulations.

\bibliographystyle{eg-alpha-doi}  
\bibliography{egbibsample}        


\end{document}